\documentclass[10pt,journal]{IEEEtran}
\usepackage{amssymb}
\usepackage{amsmath}
\usepackage{cite}
\usepackage{url}
\usepackage{hyperref}
\usepackage{xcolor}
\usepackage{cite,graphicx}
\usepackage{fancyhdr}
\usepackage{cases}
\usepackage{mdwmath}
\usepackage{mdwtab}
\usepackage{caption}
\usepackage[scr=rsfs]{mathalpha}
\usepackage{amsthm}
\usepackage{setspace}
\usepackage[linesnumbered,ruled,vlined]{algorithm2e}
\usepackage{subfig}
\usepackage{multirow}
\usepackage{makecell}
\usepackage{mathtools}

\graphicspath{{./src/img/}}

\newtheorem{theorem}{Theorem}

\newtheorem{lemma}{Lemma}

\newtheorem{corollary}{Corollary}

\allowdisplaybreaks
\title{Near-Field Sensing Enabled Predictive Beamforming: From Estimation to Tracking}

\author{Hao~Jiang,~\IEEEmembership{Graduate Student Member,~IEEE}, Zhaolin Wang,~\IEEEmembership{Graduate Student Member,~IEEE}, \\ and Yuanwei Liu,~\IEEEmembership{Fellow,~IEEE}

\thanks{

Hao Jiang and Zhaolin Wang are with the School of Electronic Engineering and Computer Science, Queen Mary University of London, London E1 4NS, U.K. (e-mail: \{hao.jiang; zhaolin.wang\}@qmul.ac.uk).

Yuanwei Liu is with the Department of Electrical and Electronic Engineering, The University of Hong Kong, Hong Kong (e-mail: yuanwei@hku.hk).
} 
}

\begin{document}
\maketitle
\begin{abstract}
A near-field sensing (NISE) enabled predictive beamforming framework is proposed to facilitate wireless communications with high-mobility channels.
Unlike conventional far-field sensing, which only captures the angle and the radial velocity of the user, NISE enables the estimation of the full motion state, including additional distance and transverse velocity information.
Two full-motion state sensing approaches are proposed based on the concepts of estimation and tracking, respectively.
1)~Adaptive gradient descent alternative optimization (AGD-AO) approach: In this approach, the full motion state of the user is estimated within a single coherent processing interval (CPI). 
In particular, the gradient descent is adopted to estimate the transverse and radial velocities of the user based on the maximum likelihood criteria, while the distance and the angle are calculated by the kinematic model.
In this process, moment estimations are leveraged to adaptively tune the step size, thereby leading to a smoother and faster gradient descent. 
2)~Extended Kalman filter (EKF) approach: In this approach, the full motion state of the user is tracked across multiple CPIs.
Based on the noisy measurements (echos) in multiple CPIs, the EKF method iteratively predicts and updates the current motion state to achieve a low tracking error.
Based on the obtained full motion state, the beam prediction, and Doppler frequency compensation can be carried out with minimum pilot overhead.
Numerical results are provided to validate the effectiveness and efficiency of the proposed approach compared to the conventional far-field predictive beamforming and feedback-based approaches. 
It is also revealed that: 1)~the proposed AGD-AO method can achieve stable descending with small gradients, thereby accelerating convergence; 2)~compared to far-field predictive beamforming and feedback-based schemes, both of the proposed methods exhibit superior performance; and 3) by incorporating multiple CPIs, the EKF method exhibits greater robustness in low signal-to-noise-ratio (SNR) regions.
\end{abstract}

\section{Introduction} \label{sect:introduction}
With the commercialization of 5G, 6G communication technology is anticipated to push the performance boundaries of 5G, by enabling a higher transmission rate of 1 Tpbs and denser connectivity of $100/\mathrm{m}^3$ \cite{chowdhury20206G, jiang2021road}.
However, the limited and saturated bandwidth of the current low-frequency band fundamentally contradicts the visions for 6G.
As a promising solution, high-frequency bands, such as the upper mid-band from 7-24 GHz, the mmWave band from 24-71 GHz, and even the sub-THz band from 92-115 GHz, can offer abundant bandwidths and enable high-speed transmissions, consequently attracting significant attention from both academia and industry \cite{bjornson2024enabling, wang2018millimeter}. 

However, transmission over high-frequency bands will inevitably suffer from severe atmospheric-induced attenuation \cite{wang2018millimeter, roh2014millimeter}.
To compensate for this attenuation, the beamforming technique is a promising solution, which can build highly directional ``pencil-like" beams to produce vast beam gains. 
To harness the benefits of beamforming, extremely large-scale antenna arrays (ELAAs) have emerged as a promising candidate for 6G technologies.
These arrays are typically equipped with several hundreds or even thousands of antennas \cite{lu20246g}.   
However, due to the narrowness of beams emitted by ELAAs, the performance of beamforming is highly dependent on the alignment between the beamforming pattern and the user's location.
Misalignment can result in significant performance loss, as reported by \cite{nitsche2015steering} that a link budget loss of around 17 dB can be caused by an 18-degree misalignment.
To address this issue while avoiding the need for channel estimation (CE) on high-dimension wireless channels, beam training or beam alignment techniques have been proposed to build robust transmission links between transceivers with low complexity prior to transmissions \cite{heng2021six}.
Nevertheless, in high-mobility communication systems, persistently maintaining such a transmission link is challenging since repeatedly sweeping through the beam training codebook can cause intolerable delays to the communication system.
To solve this problem, beam tracking is proposed, 
exploiting the temporal correlations between consecutive user locations \cite{chen2024beam, liu2022intergrated, zhou2022attitude, chuang2021adaptive}.
Therefore, beam sweeping can be done by merely sweeping a smaller subset of the entire beam training codebook, thus avoiding the need to transmit redundant pilots (code words).
Although this idea is quite simple and straightforward, it still suffers from two main drawbacks \cite{liu2022intergrated}.
First, the user needs to report feedback on its measurements of received pilots (code words) to the transmitter, thus necessitating a dedicated backhaul link.
Second, when the user is moving at high speed, the timeliness of the feedback is critical in guaranteeing the alignment between transceivers.
To mitigate the challenges of beam tracking, predictive beamforming is a more promising option for maintaining alignment without the dedicated backhaul link \cite{liu2020radar}.
Specifically, predictive beamforming can predict the motion state of the user in the subsequent coherent time interval (CPI) based on its state in the current CPI.  
Additionally, the uplink feedback in beam tracking can be eliminated and replaced by the transmitted integrated sensing and communications (ISAC) signals from the transmitter.
By doing so, the current motion state of the user can be gained via echo signals.
By leveraging the predictive beamforming technique, the timeliness of beamformers can be enhanced in the absence of a dedicated feedback link.

Although predictive beamforming has shown great potential in far-field scenarios, such as vehicular networks \cite{liu2020radar} and unmanned aerial vehicle (UAV) networks \cite{liu2021location}, the transmitter can only predict the angle and radial velocity of the user, which limits its applications in real-world scenarios.
In contrast, near-field sensing (NISE)  can capture the full motion state of the user, including angle, distance, radial velocity, and transverse velocity, based on which the predictive beamforming is enabled.
Specifically, with the implementation of ELAAs and the utilization of high-frequency bands, the enlarged antenna aperture and high carrier frequency extend the boundary of the near-field region to tens or even hundreds of meters \cite{liu2024nearfield, liu2023near}.
In the near-field region, the predominance of spherical wavefronts causes the phase variations among antenna elements to be determined jointly by angle and distance, thus giving rise to a unique distance-angle-dependent channel feature \cite{zhang2022beam}.
Empowered by this feature, NISE can facilitate the estimation of the full motion state with limited bandwidth \cite{wang2024performance}, enabling predictive beamforming to be applied to more random and complex user trajectories.
 
\subsection{Related Works}
For the conventional far-field scenarios, the predictive beamforming problem has been extensively investigated in \cite{liu2020radar, du2023integrated, yuan2021bayesian, mu2021integrated, liu2022learning, zhang2024transformer, zhang2023deep}.
The existing works on this topic can be categorized into two classes: position-based methods and channel-based methods.
Specifically, the position-based methods design the beamformer based on the user's position.
For example, the authors in \cite{liu2020radar} utilized the extended Kalman filter (EKF) method to track the users' trajectories.
By forecasting the current motion state of the user based on prior knowledge of the user's kinematic model, the future motion state of the user can be predicted, according to which the beamformer is designed.
With a similar idea of the EKF-based method, the Bayesian-based prediction methods have been exploited to reduce the computational complexity of EKF, such as the message-passing method proposed by \cite{yuan2021bayesian}.
Furthermore, authors in \cite{du2023integrated} utilized an adjustable bandwidth to detect the presence of the user.
This approach addresses the issue where the user is not covered by the pencil-sharp beams produced by massive multiple-input multiple-output (mMIMO) systems.
Due to the computational complexity of the above position-based methods, channel-based methods employ machine learning (ML) to predict the future channel directly.
For example, a deep neural network (DNN)-based method was utilized for channel prediction in \cite{mu2021integrated}.
As a further advance, some more advanced neural network structures, such as long-short term memory (LSTM) \cite{liu2022learning}, attention-based LSTM \cite{zhang2024transformer}, and Transformer \cite{zhang2023deep}, have also been studied to enhance the channel-based predictive beamforming performance in far-field systems. 
Compared to the far-field predictive beamforming techniques, research on near-field predictive beamforming is still in its infancy.
A pioneering work has been done by the authors of \cite{wang2024near}, which utilized a gradient-based method to sense the velocities of the user from the echo signals, upon which predictive beamforming was carried out.
However, since the Hessian matrix is utilized for estimation, the computational complexity of this work is very high.

\subsection{Motivations and Contributions}
By exploiting the spherical wave propagation, NISE allows for estimating the full motion state of the user in high-mobility scenarios.
In contrast, the far-field channels only allow for partial sensing of the user, including the angle and the radial velocity.
Therefore, with full motion estimation, NISE-enabled predictive beamforming can be applied to arbitrary user trajectories without prior information \cite{you2024generation}.
Although channel-based methods can be applied to the near-field scenario, these methods are limited in providing full motion state of the user. 
This limitation hampers the versatility of these methods, as the full motion state of the user can be critical for supporting other functionalities in certain applications, such as user monitoring or scheduling \cite{wang2022integrated}.
Therefore, to obtain the full motion state of the user while avoiding the need for feedback in \cite{mylonopoulos2024adaptive, chen2024beam} and high computational complexity in \cite{wang2024near},  this work presents a pair of methods for performing both user tracking and predictive beamforming.
The detailed contributions of this paper can be summarized as follows:
\begin{itemize}
    \item We study a cellular network where a multi-antenna base station (BS) serves a moving user that is adherent to an arbitrary trajectory.
    The dynamic near-field multiple-input single-output (MISO) channel is modeled and parameterized by the full motion state of the user.
    Based on the channel modeling, a predictive beamforming problem is formulated.
    
    \item
    We propose an adaptive gradient descent alternative optimization (AGD-AO) method to estimate the full motion state of the user based on the echo signals contained in one CPI. 
    Compared to the plain gradient descent adopted by \cite{wang2024near}, the AGD-AO method adopts moment estimations to overcome the small gradients around the optimum, ensuring a fast gradient decent trajectory.
    Additionally, since the calculation of Hessian in \cite{wang2024near} is exempted, the computational complexity is significantly reduced.
    
    \item
    We propose an EKF method to track the full motion state of the user by exploiting the echo signals in multiple CPIs.
    The EKF method iteratively predicts and then updates its knowledge of the current motion state, aiming to achieve lower tracking errors.
    
    \item Simulation results indicate that the proposed predictive beamforming schemes can achieve near-optimal throughput via accurately tracking the user's full motion state.
    Due to only one CPI being utilized for estimation, the performance of the proposed AGD-AO method degrades as the distance between the transceivers increases or the transmit power at the BS decreases.
    Conversely, with multiple CPIs used, the EKF method is more robust regarding the aforementioned changes.
\end{itemize}

\subsection{Organization and Notations}
The rest of the paper is arranged as follows. 
In Section \ref{sect:system model}, a NISE-enabled communication network is modeled and predictive beamforming problem is formulated.
To solve this problem, two algorithms are proposed in \ref{sect:algorithm_agd_ao} and \ref{sect:ekf_solution}.
In Section \ref{sect:simulation_res}, the simulation results are provided to validate the effectiveness of the proposed methods.
Finally, conclusions are drawn in Section \ref{sect:conclu}.

\textit{Notations:}
Scalars, vectors, and matrices are denoted by the lower-case, bold-face lower-case, and bold-face upper-case letters, respectively.
$\mathbb{C}^{M \times N}$ and $\mathbb{R}^{M \times N}$ denote the space of $M \times N$ complex and real matrices, respectively.
$(\cdot)^T$, $(\cdot)^*$, and $(\cdot)^H$ denote the transpose, conjugate, and conjugate transpose, respectively.
$\mathbf{I}_M$ and $\mathbf{0}_M$ denote the $M\times M$ identity matrix and the $M \times 1$ all-zero vector, respectively.
$|\cdot|$ represent the absolute value of a scalar.
For a matrix $\mathbf{A}$, $[\mathbf{A}]_{:,j}$, $[\mathbf{A}]_{i,:}$, and $[\mathbf{A}]_{i,j}$ denote the $j$-th column, the $i$-th row, and the $(i,j)$-th element, respectively.
For a vector $\mathbf{a}$, $[\mathbf{a}]_i$ and $\left\| \mathbf{a} \right\|_2 $ denote the $i$-th element and $2$-norm, respectively.
\section{System Model} \label{sect:system model} 
\begin{figure}[t!]
	\includegraphics[width=0.9\linewidth]{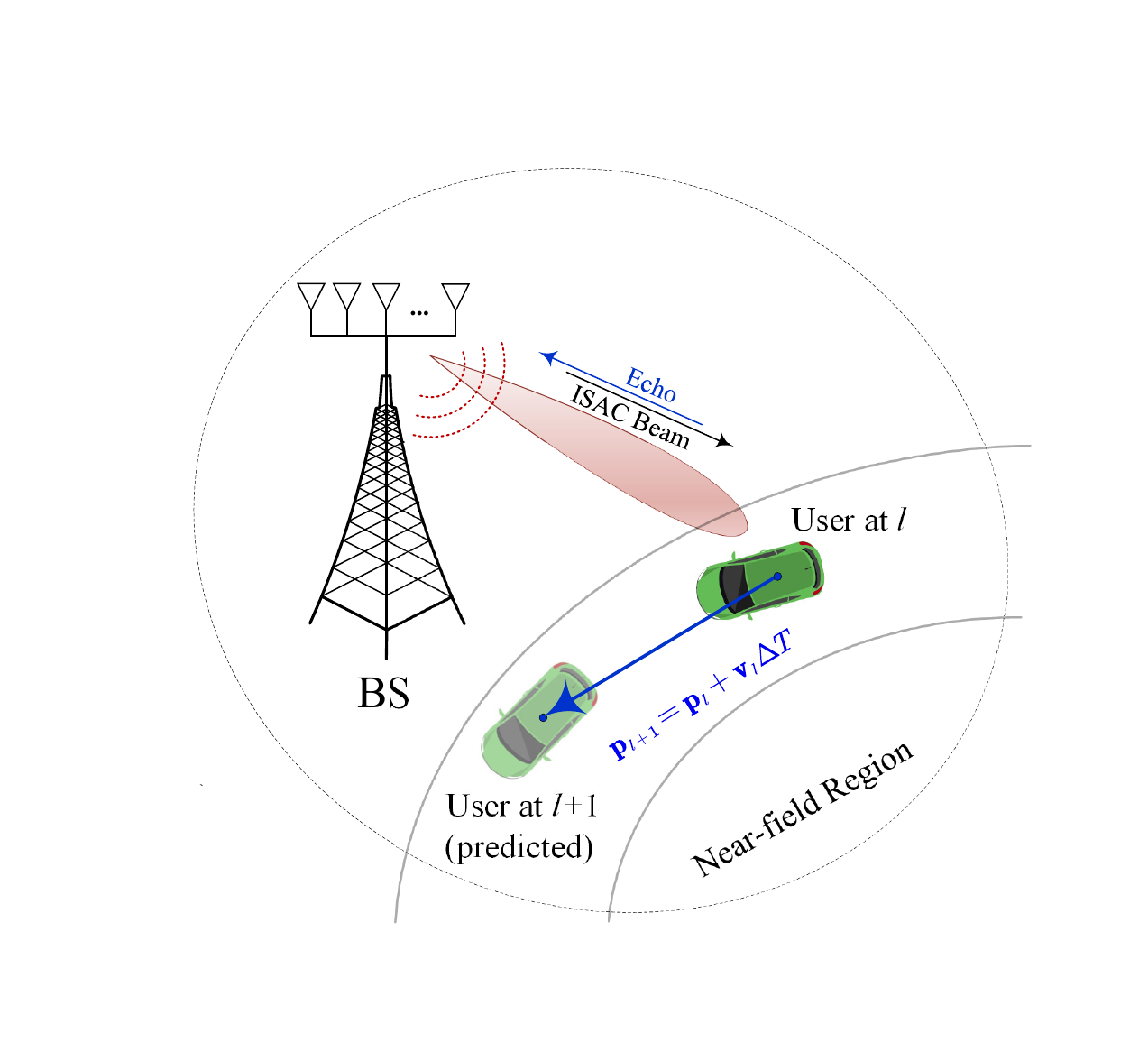}
	\caption{An overview of the system model.}
	\label{fig:system_model}
\end{figure} 
As shown in Fig. \ref{fig:system_model}, we consider a narrowband near-field communication network that includes a BS equipped with an $M$-antenna uniform linear array (ULA) and a single-antenna mobile user.
We assume that the ULA and the user are located on the $xy$ plane, with the ULA aligned along the positive direction of the $x$-axis.
It is assumed that the origin of the coordinate system is put at the center of the ULA at the BS.
The system operates in time-division duplex (TDD) mode, ensuring the reciprocity of the wireless channel.
Since high-frequency channels are dominated by the line-of-sight (LoS) link, we thus ignore the impact caused by the non-line-of-sight links \cite{liu2020radar}. 
In addition, we assume that the position of the user is beyond the reactive near-field region that is confined to just a few wavelengths from the antenna array \cite{ouyang2024impact}.
Letting the symbol duration be $T_s$, $N$ symbol durations are grouped into one CPI.
Each CPI lasts $\Delta T=NT_s$ and is indexed by $l$.
Owing to the short duration of each CPI, we assume that the motion state of the user remains unchanged within one CPI.
In addition, the BS adopts full-duplex antenna architecture detailed in \cite{smida2023full}, wherein the circulators are utilized to enable simultaneous transmission and reception.
Such a full-duplex technique is adopted to eliminate the self-interference.
A fully digital antenna configuration is considered in this work. 

\subsection{Near-Field Channel Model}
\begin{figure}[t!]
    \includegraphics[width=1\linewidth]{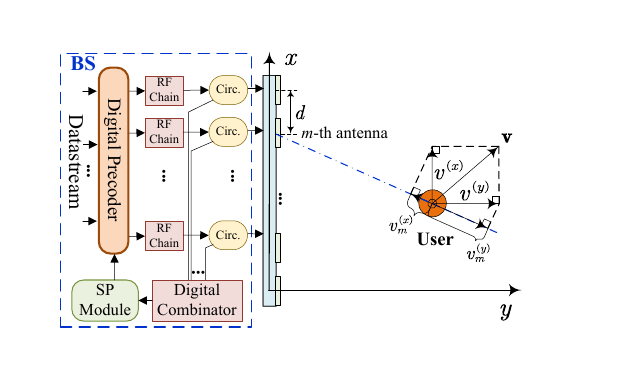}
	\caption{An overview of the antenna geometry.}
	\label{fig:antenna_geometry}
\end{figure}
Let $\mathbf{p}=\left[ x,y \right] ^T \in \mathbb{R}^{2 \times 1}$  and $\mathbf{v} \triangleq [v^{(x)}, v^{(y)}]^T \in \mathbb{R}^{2 \times 1}$ denote the position and velocity of the user within a specific CPI, respectively, as depicted in Fig. \ref{fig:antenna_geometry}.
In particular, $v^{(x)}$ and $v^{(y)}$ represent the velocity components in the positive direction of $x$-axis and $y$-axis, respectively.
The position and velocity vectors can be related to the angle, distance, radial velocity, and transverse velocity of the user with respect to the center of the ULA as follows:
\begin{align}
\begin{cases}
    \theta =\mathrm{arc}\tan \left( \frac{y}{x} \right), \\
r=\sqrt{x^{2}+y^{2}}, \\
v^{\left( r \right)}=v^{\left( x \right)}\cos \theta+v^{\left( y \right)}\sin \theta, \\
v^{\left( \theta \right)}=v^{\left( x \right)}\sin \theta+v^{\left( y \right)}\cos \theta.
\end{cases}
\end{align}
In far-field systems, only angle $\theta$ and radial velocity $v^{\left( r \right)}$ can be estimated due to the planar-wave propagation.
On the contrary, NISE can estimate $r$ and $v^{\left( \theta \right)}$ in addition to $\theta$ and $v^{\left( r \right)}$, thereby leading to the capture of full motion state of the user \cite{wang2024near}. 
In this study, we utilize the $xy$-based vectors to streamline the algorithm design.
Let $d$ denote the antenna spacing of the ULA. 
Then, the coordinates of the $m$-th antenna can be specified by $\mathbf{k}_{m}=[\delta_m d ,0 ]^T$, where $\delta_m = m-1-\frac{M-1}{2}$, $m\in \mathcal{M}$, and $\mathcal{M} \triangleq \{1, 2, ..., M\}$.
In the near-field region, the distance between the $m$-th antenna and the user must be calculated as the Euclidean distance as
\begin{align}
	r_{m}^{}\left( \mathbf{p} \right) =\left\| \mathbf{k}_m - \mathbf{p} \right\| _{2}^{}\in \mathbb{R}, \qquad \forall~m \in \mathcal{M}.
\end{align}
Therefore, according to \cite{liu2023near} and \cite{lu2024tutorial}, the stationary near-field array response vector with respect to the user's position can be written as 
\begin{align}
    \tilde{\mathbf{a}}(\mathbf{p}) = \left[ e^{{-j \frac{2\pi}{\lambda} r_{1}^{}\left( \mathbf{p} \right)}},...,{}e^{{-j\frac{2\pi}{\lambda}r_{M}^{}\left(\mathbf{p} \right)}} \right] ^T. \label{eq:static_array_response}
\end{align}
In high-mobility scenarios, the Doppler frequency caused by the velocity of the user also needs to be taken into account. \cite{tse2005fundamentals}.
Due to the spherical wave propagation in the near-field region, the velocity projections with respect to different antennas are not uniform, thus giving rise to non-uniform Doppler frequencies.
To quantify the Doppler frequencies, according to the geometry illustrated in Fig. \ref{fig:antenna_geometry}, the projection of $v^{(x)}$ and $v^{(y)}$ onto the line-of-sight direction between the user and the $m$-th antenna can be calculated as
\begin{align}
	v_{m}^{\left( x \right)}&=g_{m}\left( \mathbf{p} \right) v^{\left( x \right)},
	\\
	v_{m}^{\left( y \right)}&=q_{m}\left( \mathbf{p} \right) v^{\left( y \right)},
\end{align}
where $g_{m}\left( \mathbf{p} \right)$ and $q_{m}\left( \mathbf{p} \right)$ denote the projection coefficients and are given by 
\begin{align}
	g_{m}\left( \mathbf{p} \right) \triangleq \frac{\left| [\mathbf{k}_m]_1 - x \right|}{r_{m}^{}\left( \mathbf{p} \right)}, \qquad
	q_{m}\left( \mathbf{p} \right) \triangleq \frac{\left| [\mathbf{k}_m]_2 - y \right|}{r_{m}^{}\left( \mathbf{p} \right)}. \label{eq:q}
\end{align}
Given that $v_{m}^{\left( x \right)}$ and $v_{m}^{\left( y \right)}$ are collinear, the composite velocity with respect to the $m$-th antenna can be presented by
\begin{align}
	v_{m}^{}=v_{m}^{\left( x \right)} + v_{m}^{\left( y \right)}. \label{eq:composite_velocity}
\end{align}
With the above element-wise velocity decomposition, the Doppler-frequency vector can be expressed as
\begin{align}
	\mathbf{d}\left( n; \mathbf{v} \right) =\left[ e^{-j\frac{2\pi}{\lambda}n T_s v_{1}^{}},...,e^{-j\frac{2\pi}{\lambda}nT_sv_{M}^{}} \right] ^T, \label{eq:dynamic_array_response}
\end{align}
where $n \in \mathcal{N}$ denotes the index of symbols, and $\mathcal{N} \triangleq \{1,2, ..., N\}$.
Since the user is located beyond the reactive near-field region, the amplitude variations among antennas can be neglected, thus resulting in a uniform channel gain \cite{Emil2021primer}.
Without loss of generality, the channel gain is calculated based on the central link, i.e., the distance between the coordinate origin and the user.
Therefore, the channel gain during one CPI can be evaluated by $\alpha_1(\mathbf{p})$, which is a function of the user's position $\mathbf{p}$.
Jointly considering pathloss $\alpha_1(\mathbf{p})$, stationary array response vector $\tilde{\mathbf{a}}(\mathbf{p})$, and Doppler-frequency vector $\mathbf{d}\left( n; \mathbf{v} \right)$, the overall downlink near-field channel vector $\mathbf{h}(n; \mathbf{v},\mathbf{p}) \in \mathbb{C} ^{M\times 1} $ at the $n$-th symbol duration can be written as
\begin{align}
    \mathbf{h}(n; \mathbf{v},\mathbf{p})=\alpha _1(\mathbf{p})\tilde{\mathbf{a}}(\mathbf{p})\odot \mathbf{d}\left( n;\mathbf{v} \right) =\alpha _1(\mathbf{p})\mathbf{a}(n; \mathbf{v},\mathbf{p}),\label{eq:BS_veh_channel}
\end{align}
where $\mathbf{a}(n; \mathbf{v},\mathbf{p})\triangleq \tilde{\mathbf{a}}(\mathbf{p})\odot \mathbf{d}\left( n;\mathbf{v} \right)$ denotes the overall array response vector for a moving user.
To obtain the position of the user, the BS can gather echo signals that the user reflects. 
Hence, the round-trip channel, i.e., from BS to the user and then back to BS, needs to be modeled. 
For attenuation, the channel gain of the round-trip channel is denoted by $\alpha _2(\mathbf{p})$, which contains the impact of the radar cross section (RCS), pathloss, and array gains.
For the attenuation-free array response matrix, we can exploit the reciprocity of the wireless channel by $\mathbf{a}(n;\mathbf{v},\mathbf{p})\mathbf{a}^T(n;\mathbf{v},\mathbf{p})$.
Then, by jointly considering the attenuation and the array response matrix, the round-trip channel matrix can be presented by the following equation: 
\begin{align}
	\mathbf{H}\left( n;\mathbf{v},\mathbf{p} \right) =\alpha _2(\mathbf{p}) \mathbf{a}(n;\mathbf{v},\mathbf{p})\mathbf{a}^T(n;\mathbf{v},\mathbf{p}).\label{eq:BS_BS}
\end{align} 
Here, we assume that the RCS follows the Swerling I model.

\subsection{Signal Model}
In this subsection, we will present the communication and the echo signal model, respectively.
In particular, the transmit signal during the $l$-th CPI from the BS can be modeled as
\begin{align}
	\mathbf{x}_l(n) = \mathbf{f}_l(n) s_l(n),
\end{align}
where $s_l(n)$ denotes data symbols and $\mathbf{f}_l(n)$ denotes the beamforming vector. 
We consider an average power constraint on the data symbols, i.e., $\mathbb{E}\{|x_l(n)|^2\}=P$, with $P$ denoting the average transmit power.
Then, based on the channel model in \eqref{eq:BS_veh_channel}, the received communication signal at the user can be represented as
\begin{align}
	y_{l}(n)=\mathbf{h}^{T}(n;\mathbf{v}_l,\mathbf{p}_l)\mathbf{x}_l(n)+z_l(n), \label{eq:received_signal}
\end{align}
where $\mathbf{v}_l$ and $\mathbf{p}_l$ denote the velocity and position vectors of the user during the $l$-th CPI and $z_l(n) \sim \mathcal{CN}(0, \sigma^2_{c})$ denotes the complex Gaussian noise with power $\sigma_{c}^2$.
Similarly, based on the channel model in \eqref{eq:BS_BS}, the received echo signal at the BS can be written as 
\begin{align}
    \tilde{\mathbf{y}}_l(n;\mathbf{v}_l,\mathbf{p}_l) &=\mathbf{H}(n;\mathbf{v}_l,\mathbf{p}_l)\mathbf{x}_l(n)+\tilde{\mathbf{z}}_{l}(n), \label{eq:sensing_signal}
\end{align}
where $\tilde{\mathbf{z}}_{l}(n) \sim \mathcal{CN}(\boldsymbol{0}_M, \sigma_e^2\mathbf{I}_{M})$ denotes the complex Gaussian noise at the BS with $\sigma_e^2$ being the noise power.
It is noted that the intended signal $s_l(n)$ is known at the BS.
Therefore, sufficient statistics (observation vector) for estimating the motion state of the user can be obtained using the matched filtering as \cite{bekkerman2006target}:
\begin{align}	&\mathbf{y}_l(n;\mathbf{v}_l,\mathbf{p}_l)=\tilde{\mathbf{y}}_l (n;\mathbf{v}_l,\mathbf{p}_l)s_{l}^{*}(n) \notag \\
&=\mathbf{H}\left(  n;\mathbf{v}_l,\mathbf{p}_l \right) \mathbf{f}_l(n)+\tilde{\mathbf{z}}_l(n)s_{l}^{*}(n) \triangleq h\left( n;\mathbf{v}_l,\mathbf{p}_l \right) +\mathbf{z}_l^{(1)}(n)
	, \label{eq:observation}
\end{align}
where $h(n;\mathbf{v},\mathbf{p})$ is the observation function.
In addition, since the intended signal $s_{l}(n)$ and the noise term $\mathbf{z}_l^{(1)}(n)$ are independent, the observation noise $\mathbf{z}_l^{(1)}(n) \sim \mathcal{CN}(\boldsymbol{0}_M, \mathbf{R})$ holds, where $\mathbf{R} \triangleq \sigma_e^2 \mathbf{I}_{M}$ denotes the covariance matrix of the observation noise. 

\subsection{Transmission Protocol and Motion State Model} \label{sect:transmission_protocol_mobility_model}
\begin{figure}[t]
	\centering
	\includegraphics[width=1\linewidth]{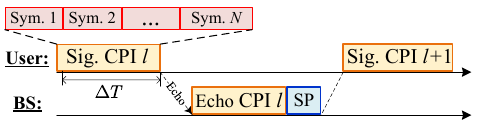}
	\caption{An illustration of the transmission protocol, where ``SP" refers to signal processing, ``Sym." refers to symbol, and ``Sig." refers to signal.}
	\label{fig:transmission_protocol}
\end{figure}
The transmission protocol is illustrated in Fig. \ref{fig:transmission_protocol}, composed by the following three steps: 1)~the BS transmit $N$ symbols during the $l$-th CPI using beamformers $\mathcal{F}_l =\{\mathbf{f}_l(1), ..., \mathbf{f}_l(N)\}$; 2)~the BS receives echo signals to form the observations $\mathcal{Y}_l = \{\mathbf{y}_{l}(1), ..., \mathbf{y}_{l}(1)\}$; and 3) the user's motion state is extracted from $\mathcal{Y}_l$ during the signal processing block to construct beamformers $\mathcal{F}_{l+1}$ to facilitate the transmissions in the subsequent $(l+1)$-th CPI.
Based on the above transmission protocol, the motion state model of the user is specified as:
\begin{align}\label{eq:mobility-xy}
	\begin{cases}
		x_{l+1}=x_{l}+v_{l}^{(x)}\Delta T, \\ 
		y_{l+1}=y_{l}+v_{l}^{(y)}\Delta T, \\ 
	\end{cases}
\end{align} 
where the subscript $l$ represents that the quantity is for the $l$-th CPI. 
Additionally, we also consider the velocity variance within the $l$-th CPI, which is depicted by 
\begin{align}\label{eq:mobility-v}
	\begin{cases}
		v_{l+1}^{(x)}=v_{l}^{(x)}+\Delta v_{l}^{(x)}, \\
		v_{l+1}^{(y)}=v_{l}^{(y)}+\Delta v_{l}^{(y)}, \\
	\end{cases}
\end{align}
where $\Delta v_{l}^{(x)} \sim \mathcal{N}(0, \sigma_{v^{(x)}}^2)$ and $\Delta v_{l}^{(y)}\sim \mathcal{N}(0, \sigma_{v^{(y)}}^2)$ are the velocity variances following independent Gaussian distributions.
It is important to highlight that due to the short duration of each CPI, the variances of velocities between consecutive CPIs are typically very small.
More compactly, the motion state can be written as
\begin{align}
	\boldsymbol{\eta }_{l+1}=f\left( \boldsymbol{\eta }_{l} \right) +\mathbf{z}_l^{(2)}, \label{eq:process}
\end{align} 
where $\boldsymbol{\eta }_{l} \triangleq [x_l, y_l, v^{(x)}_l, v^{(y)}_l]^T$ is the motion state vector, and $f(\cdot): \mathbb{R}^{4 \times 1} \mapsto \mathbb{R}^{4 \times 1}$ is the kinematic function, which is defined by 
\begin{align}
    f\left( \boldsymbol{\eta }_{l} \right) =\left[ \begin{matrix}
	1&		0&		\bigtriangleup T&		0\\
	0&		1&		0&		\bigtriangleup T\\
	0&		0&		1&		0\\
	0&		0&		0&		1\\
    \end{matrix} \right] \boldsymbol{\eta }_{l},  \label{eq:J_f}
\end{align}
 and $\mathbf{z}_l^{(2)} \sim \mathcal{N}(\mathbf{0}_4, \mathbf{Q})$ denotes the motion noise with a covariance matrix defined by $\mathbf{Q} \triangleq \mathrm{diag}\{0, 0, \sigma_{v^{(x)}}^2, \sigma_{v^{(y)}}^2\} \in \mathbb{R}^{4 \times 4}$.

\subsection{Problem Formulation}
Based on \eqref{eq:received_signal}, the received signal-to-noise ratio (SNR) during the $n$-th symbol time of the $(l+1)$-th CPI can be written as 
\begin{align}
	\gamma _{l+1}^{}\left( n \right) =\frac{P\left| \mathbf{h}^T (n;\mathbf{v}_{l+1},\mathbf{p}_{l+1})\mathbf{f}_{l+1}(n) \right|^2}{\sigma_{c}^2}. \label{eq:snr}
\end{align}
Therefore, the average throughput during the $(l+1)$-th CPI can be quantified as 
\begin{align}
	R_{l+1}^{}\left( \mathcal{F} _{l+1} \right) =\frac{1}{N}\sum_{n=1}^N{\log _2\left( 1+\gamma _{l+1}^{}\left( n \right) \right)}. \label{eq:throughput}
\end{align}
If the channel is perfectly known, the beamformer can be optimally designed according to the concept of maximum radio transmission (MRT). 
But this requires extremely high channel estimation overheads, especially for the considered time-variant channel model. Fortunately, with $\mathcal{Y}_l$ at hand. it is possible to achieve near-optimal beamforming design, indicating that the beamformers need to be designed in a predictive manner.
Therefore, a mapping function $U(\cdot)$ needs to be found to map the observations $\mathcal{Y} _{l}$ to the beamformers $\mathcal{F} _{l+1}$, such that the average throughput in \eqref{eq:throughput} can be maximized.
Therefore, the predictive beamforming problem can be formulated as follows:
\begin{subequations}
	\begin{align} 
	&\max_{\mathcal{F} _{l+1}=U\left( \mathcal{Y} _{l} \right)} R_{l+1}^{}\left( \mathcal{F} _{l+1} \right), \label{eq:obj}\\
		&~~~~{\rm s.t.}~\left\| \mathbf{f}_{l+1}\left( n \right) \right\| _{2}^{2}=1, ~~\mathrm{for}~ \forall \,\,\mathbf{f}_{l+1}\left( n \right) \in \mathcal{F} _{l+1}, \label{eq:constraint}
	\end{align}
\end{subequations}
where \eqref{eq:constraint} represents the unit-power constraint introduced by the fully digital antenna configuration.
In the sequel, we will present how to solve \eqref{eq:obj} based on the observations in the former CPI, i.e., $\mathcal{Y} _{l}$.

According to \eqref{eq:snr} and \eqref{eq:throughput}, the closed-form solution for \eqref{eq:obj} at the $n$-th symbol time of the $(l+1)$-th CPI can be calculated as follows:
\begin{align}
    \mathbf{f}_{l+1}^{\rm opt}\left( n \right) &=\frac{1}{\sqrt{M}}\mathbf{a}^*(n;\mathbf{v}_{l+1},\mathbf{p}_{l+1}) \notag \\
    &=\frac{1}{\sqrt{M}}\tilde{\mathbf{a}}^*(\mathbf{p}_{l+1})\odot \mathbf{d}^*\left( n;\mathbf{v}_{l+1} \right), 
\label{eq:solution}
\end{align}
which is parameterized by the position $\mathbf{p}$ and the velocity $\mathbf{v}$ during $(l+1)$-th CPI, denoted by $\mathbf{p}_{l+1}$ and $\mathbf{v}_{l+1}$, respectively.
According to \eqref{eq:BS_BS} and \eqref{eq:observation}, the observations during the current CPI $\mathcal{Y} _{l}$ contains the position and velocity of the $l$-th CPI, denoted by $\mathbf{p}_{l}$ and $\mathbf{v}_{l}$, respectively.
Therefore, if $\mathbf{p}_{l}$ and $\mathbf{v}_{l}$ can be estimated, the predicted position $\tilde{\mathbf{p}}_{l+1}$ during the $(l+1)$-th CPI can be calculated by 
\begin{align}
    \tilde{\mathbf{p}}_{l+1}=\hat{\mathbf{p}}_{l}+\Delta T \hat{\mathbf{v}}_{l}, \label{eq:positional}
\end{align}
where $\hat{\mathbf{p}}_{l}$ and $\hat{\mathbf{v}}_{l}$ are estimations of $\mathbf{p}_{l}$ and $\mathbf{v}_{l}$ of the $l$-th CPI.
In contrast, velocity $\tilde{\mathbf{v}}_{l+1}$ cannot be directly obtained due to the unknown acceleration between CPIs as outlined by \eqref{eq:mobility-v}.
Nevertheless, due to the small accelerations and short durations of CPIs, it is practical to utilize the estimated velocity in the current CPI to approximate that in the subsequent CPI  \cite{liu2020radar, wang2024near}, which is also known as Doppler frequency compensation, i.e.,    
\begin{align}
\tilde{\mathbf{v}}_{l+1} \simeq \hat{\mathbf{v}}_{l}. \label{eq:velocity}
\end{align}
Based on \eqref{eq:positional} and \eqref{eq:velocity}, the predictive beamformers $\mathcal{F} _{l+1}$ can be designed according to
\begin{align}
	\mathbf{f}_{l+1}^{\rm opt}(n)\simeq \frac{1}{\sqrt{M}}\tilde{\mathbf{a}}^*(\tilde{\mathbf{p}}_{l+1})\odot \mathbf{d}^*\left( n;\tilde{\mathbf{v}}_{l+1} \right), ~\mathrm{for}~\forall n\in \mathcal{N} \label{eq:beamformer}
\end{align}
The above procedure is summarized by \textbf{Algorithm \ref{alg:beamformer_design}}.

\begin{algorithm}[t!]
    \SetAlgoLined 
	\caption{Position-Based Predictive Beamforming}\label{alg:beamformer_design}
	\KwIn{Estimated user's position and velocity for the current $l$-th CPI, i.e., $\hat{\mathbf{p}}_{l}$ and $\hat{\mathbf{v}}_{l}$. } 
	\KwOut{Beamformers for the next CPI, i.e., $\mathcal{F} _{l+1}$ }
	\tcp{Obtain the Positional Beamformer:}
    Obtain user's position for the $(l+1)$-th CPI, i.e., $\tilde{\mathbf{p}}_{l+1}$, via \eqref{eq:positional} and calculate the positional beamforming vector $\tilde{\mathbf{a}}(\mathbf{p}_{l+1})$ using \eqref{eq:static_array_response}\;
	\ForEach{index of symbol $n$}{
          \tcp{Perform Doppler Compensation:} \
          Design the Doppler compensation vector $\mathbf{d}\left( n,\tilde{\mathbf{v}}_{l+1} \right)$ of the next CPI using the approximation in \eqref{eq:velocity}\;
          \tcp{Design Predictive Beamformer:} \
          Calculate $\mathbf{f}_{l+1}^{\rm opt}(n)$ with $\tilde{\mathbf{a}}(\mathbf{p}_{l+1})$ and $\mathbf{d}\left( n;\mathbf{v}_{l+1} \right)$ according to \eqref{eq:beamformer}\;
          Move to the next symbol by $n=n+1$\;
	}
	\Return{Beamformers for the next CPI, i.e., $\mathcal{F} _{l+1}$.}
\end{algorithm}

It is indicated by the above discussions that the crux of solving \eqref{eq:obj} is to obtain $\mathbf{p}_{l}$ and $\mathbf{v}_{l}$ from $\mathcal{Y} _{l}$.
In addition, initial position and velocity $\mathbf{p}_{1}$ and $\mathbf{v}_{1}$ is obtained by initial access (IA), which is specified by 5G new radio (NR) beam management framework \cite{heng2024grid}.
Here, we utilize the $N$-th observation $\mathbf{y}_{l}(N)$ to obtain $\mathbf{p}_{l}$ and $\mathbf{v}_{l}$.
The reasons are two-fold: 1)~as $N$ increases, $\mathcal{Y}_{l}$ will have a larger cardinality, which will burden the signal process step; and 2)~since $T_s$ is small, the Doppler frequencies contained in observations with smaller index are not prominent. 
Besides, for notational brevity, $h(N;\mathbf{v}_l,\mathbf{p}_l)$, $\mathbf{H}\left( N;\mathbf{v}_l,\mathbf{p}_l \right) 
$, $\mathbf{f}_l(N)$, and $\mathbf{y}_l(N)$ are denoted as $h(\mathbf{v}_l,\mathbf{p}_l)$, $\mathbf{H}\left( \mathbf{v}_l,\mathbf{p}_l \right) 
$, $\mathbf{f}_l$, and $\mathbf{y}_l$, respectively.

In the following, we propose two methods to solve problem \eqref{eq:obj}, i.e., AGD-AO and EKF, respectively. 
The differences between the algorithms are summarized as follows:
\begin{itemize}
	\item \textbf{AGD-AO:} This method aims at estimating the velocity of the user with the observation in single CPI, while the position of the user can be calculated according to the kinematic function. 
    \item \textbf{EKF:} This method aims at tracking both the position and the velocity of the user across multiple CPIs.
\end{itemize}

\section{AGD-AO Method for Predictive Beamforming} 
\label{sect:algorithm_agd_ao}
Inspired by the method proposed in \cite{wang2024near}, we propose an AGD-AO method for predictive beamforming.
Specifically, we utilize a gradient-based method to estimate the user's velocity according to the maximum likelihood (ML) function.
Specifically, we employ a gradient-based method to decrease the value of the ML criteria in order to find the estimated velocities that are most likely to be the true velocities.
Then, the user's position is calculated according to the kinematic function. 
However, the challenge for doing so is two-fold: 1)~due to the short duration of the CPI, the Doppler frequencies contained in $\mathbf{y}_{l}$ are very small, thus leading to small gradients around the true velocity (optimization objective); and 2)~due to the presence of accelerations, the optimization objective is non-static. 
Therefore, we leverage an adaptive gradient descent method to accelerate the convergence of the algorithm.
Moreover, since we model the channel within the Cartesian coordinate system rather than the polar coordinate system, the gradients with respect to $v^{(x)}$ and $v^{(y)}$ are in the same order, thus facilitating steady gradient decent on the optimization landscape. 

\subsection{ML Estimator and Its Gradient}
According to the ML rule, we are aiming at finding the $\boldsymbol{\eta}_l$, such that the calculated observation is most similar to the real observation.
Here, the likelihood function is defined based on 2-norm, i.e., $\left\| \mathbf{y}_l-h(\boldsymbol{\eta}_l) \right\| _{2}^{2}$, where we refer $h(\mathbf{v}_l,\mathbf{p}_l)$ as $h(\boldsymbol{\eta}_l)$.
In order to increase the similarity or equivalently decrease the value of the objective function, the first step is to find the gradient of the objective function with respect to $\boldsymbol{\eta}$.
The objective function can be simplified via the following steps:
\begin{align}
    \hat{\boldsymbol{\eta }}_l&=\underset{\boldsymbol{\eta }}{\mathrm{arg}\min}\,\,\left\| \mathbf{y}_l-h(\boldsymbol{\eta}) \right\| _{2}^{2} \notag \\
    &=\underset{\boldsymbol{\eta }}{\mathrm{arg}\min}\,\,\left\| \mathbf{y}_l- \mathbf{H}\left( \boldsymbol{\eta} \right) \mathbf{f}_l \right\| _{2}^{2} \notag \\
    &=\underset{\boldsymbol{\eta }}{\mathrm{arg}\min}\,\,\left\| \mathbf{y}_l \right\| _{2}^{2}-2\Re \left\{ \mathbf{y}_l^{H}{\mathbf{H}}\left( \boldsymbol{\eta } \right) \mathbf{f}_l \right\} + \left\| {\mathbf{H}}\left( \boldsymbol{\eta } \right) \mathbf{f}_l \right\| _{2}^{2} \notag \\ 
    &=\underset{\boldsymbol{\eta }}{\mathrm{arg}\max}\,\,2\Re \left\{ \mathbf{y}_{l}^{H}{\mathbf{H}}\left( \boldsymbol{\eta } \right) \mathbf{f}_l \right\} -\left\| {\mathbf{H}}\left( \boldsymbol{\eta } \right) \mathbf{f}_l \right\| _{2}^{2}.
\label{eq:obj-AGO-AO}
\end{align}
Here, we define 
\begin{align}
    &g\left( \mathbf{b}\left( \boldsymbol{\eta } \right) \right) = \notag \\ &\qquad \quad \underset{\boldsymbol{\eta }}{\mathrm{arg}\max}\,\,2\Re \left\{ \mathbf{y}_{l}^{H}{\mathbf{H}}\left( \boldsymbol{\eta } \right) \mathbf{f}_l \right\} -\left\|{\mathbf{H}}\left( \boldsymbol{\eta } \right) \mathbf{f}_l \right\| _{2}^{2}, \label{eq:obj-AGD-AO}
\end{align}
where $\mathbf{b}\left( \boldsymbol{\eta } \right) \triangleq {\mathbf{H}}\left( \boldsymbol{\eta } \right) \mathbf{f}_l$.
To maximize \eqref{eq:obj-AGD-AO}, $\frac{\partial g\left( \mathbf{b}\left( \boldsymbol{\eta } \right) \right)}{\partial \boldsymbol{\eta }}$ needs to be derived, upon which gradient ascent method can be utilized. 
Note that $g(\cdot)$ is a real-valued scalar function of $(\cdot)$ and $\boldsymbol{\eta}$ is a real-valued vector.
Therefore, the chain rule for finding the derivative of a complex-valued composite function can be further simplified.
Therefore, $\frac{\partial g\left( \mathbf{b}\left( \boldsymbol{\eta } \right) \right)}{\partial \boldsymbol{\eta }}$ can be expressed as follows:
\begin{align}
    \nabla _{\boldsymbol{\eta }}g\left( \mathbf{b}\left( \boldsymbol{\eta } \right) \right) =2\Re \left\{ \frac{\partial g\left( \mathbf{b}\left( \boldsymbol{\eta } \right) \right)}{\partial \mathbf{b}^{T}\left( \boldsymbol{\eta } \right)}\frac{\partial \mathbf{b}\left( \boldsymbol{\eta } \right)}{\partial \boldsymbol{\eta }} \right\}.\label{eq:grad}
\end{align}
According to \eqref{eq:grad}, both the position and velocity estimations need to be done, which can introduce high computational complexity.
To mitigate this issue, we notice that the position in $l$-th CPI can be obtained by $\boldsymbol{\eta}_{l-1}$ by using the kinematic function.
Hence, in the $l$-th CPI, we can only estimate the velocity of the user.
Specifically, the estimated position $\hat{\mathbf{p}}_{l}$ at the $l$-th CPI can be calculated according to $\hat{\mathbf{p}}_{l} = \hat{\mathbf{p}}_{l-1}+\Delta T\hat{\mathbf{v}}_{l-1}
$.
Therefore, we denote $\mathbf{b}\left( \boldsymbol{\eta } \right) |_{\hat{\mathbf{p}}_{l}}=\mathbf{b}\left( {\mathbf{v}} \right)$, indicating that $\mathbf{b}(\cdot)$ is now parameterized by the estimated position $\hat{\mathbf{p}}_{l}$, with the velocity ${\mathbf{v}}$ being the only unknown variable.
Correspondingly, since only the gradient with respect to velocity is of our interest, the optimization \eqref{eq:obj-AGD-AO} can be rewritten as $\hat{\mathbf{v}}_{l}^{}=\mathrm{arg}\max _{\mathbf{v}}\,g\left( \mathbf{b}\left( \mathbf{v} \right) \right)$.
Besides, for the gradient calculation, $\nabla _{\boldsymbol{\eta }}g\left( \mathbf{b}\left( \boldsymbol{\eta } \right) \right)$ in \eqref{eq:grad} can be recasted as $\nabla _{\mathbf{v}}g\left( \mathbf{b}\left( \mathbf{v} \right) \right) $.
Therefore, we can calculate the gradient with respect to the velocity with the following equations:
\begin{align}
    \frac{\partial g\left( \mathbf{b}\left( \mathbf{v} \right) \right)}{\partial \mathbf{b}^T\left( \mathbf{v} \right)} &= \mathbf{y}_{l}^{H}-\mathbf{b}^{H}\left( \mathbf{v} \right) , \label{eq:first_partial_derivitive}\\
 \frac{\partial \mathbf{b}^{}\left( \mathbf{v} \right)}{\partial \mathbf{v}}&=-\frac{2\pi j\bigtriangleup T}{\lambda}\mathbf{A}(\hat{\mathbf{p}}_l)\odot \left( \mathbf{g}_l\odot \mathbf{d}\left( \mathbf{v} \right) \mathbf{d}^T\left( \mathbf{v} \right) \right) \mathbf{f}_l
, \label{eq:second_partial_derivitive}
\end{align}
where we have 
\begin{align}
    &\mathbf{A}(\hat{\mathbf{p}}_{l})\triangleq \alpha^{(2)}(\hat{\mathbf{p}}_{l})\tilde{\mathbf{a}}(\hat{\mathbf{p}}_{l})\tilde{\mathbf{a}}^{T}(\hat{\mathbf{p}}_{l}), \\
    &\mathbf{g}_l=\begin{cases}
	\left[ g_{1}(\hat{\mathbf{p}}_{l}),g_{2}(\hat{\mathbf{p}}_{l}),...,g_{M}(\hat{\mathbf{p}}_{l}) \right] ^T, \quad\mathrm{for}~ v^{\left( x \right)},\\
	\left[  q_{1}(\hat{\mathbf{p}}_{l}),q_{2}(\hat{\mathbf{p}}_{l}),...,q_{M}(\hat{\mathbf{p}}_{l}) \right] ^T, \quad\mathrm{for}~ v^{\left( y \right)}.\\
\end{cases}
\end{align}

\subsection{Adaptive Gradient-Based Method}
\begin{figure}
	\centering
	\includegraphics[width=0.6
 \linewidth]{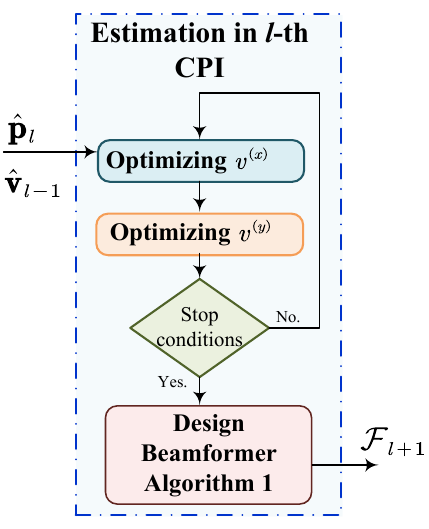} 
	\caption{An illustration of the AGD-AO method.} \label{fig:AGD-AO}
\end{figure}

To overcome the small gradients and the non-static objective, we resort to the adaptive moment estimation (Adam) method \cite{kingma2017adam}, which was first invented to solve gradient-based optimization problems.
Particularly, by integrating the adaptive gradients (AdaGrad) and the root mean squared propagation (RMSProp), Adam is capable of adaptive estimating lower order of moments (mean and uncentered variance), thus leading to a smoother gradient descent trajectory.
In addition, to stabilize the optimization process, we adopt an alternative optimization (AO) framework for Adam, meaning that $v^{(x)}$ and $v^{(y)}$ are alternatively optimized.
To do so, the hyperparameters for $v^{(x)}$ are denoted by $\zeta _x \in [0, 1)$, $\varpi _x \in [0, 1)$, and $\alpha _x$, where $\zeta _x$ and $\varpi _x$ are utilized to control the exponential decay rate of moving average, and $\alpha _x$ represents the step size on the search direction. 
Correspondingly, the hyperparameters for $v^{(y)}$ are denoted by $\zeta _y \in [0, 1)$, $\varpi _y \in [0, 1)$, and $\alpha _y$, respectively.
The optimization procedure for $v^{(x)}$ and $v^{(y)}$ contains four steps, i.e., 1)~calculating the gradients, 2)~updating biased moment estimations, 3)~computation of bias corrections, and 4)~updating parameters.
Given that the current iteration index is $k$, the partial derivatives with respect to $v^{(x)}$ and $v^{(y)}$ are respectively denoted by
\begin{subnumcases}
    ~g_{x}^{\left( k \right)}\triangleq \nabla _{v^{(x)}}g\left( \mathbf{b}\left( \mathbf{v} \right) \right) \mid _{v_{k-1}^{(y)}}, \label{eq:grad_v_x}\\
    g_{y}^{\left( k \right)}\triangleq \nabla _{v^{(y)}}g\left( \mathbf{b}\left( \mathbf{v} \right) \right) \mid _{v_{k}^{(x)}}, \label{eq:grad_v_y}
\end{subnumcases}
where $v_{k-1}^{(y)}$ and $v_{k}^{(x)}$ denotes $v^{(y)}$ and $v^{(x)}$ of the $(k-1)$-th and $k$-th iteration, respectively.
Then the first step is given by
\begin{subnumcases}
	~m_{x}^{\left( k \right)}=\zeta _{x}m_{x}^{\left( k-1 \right)}+\left( 1-\zeta _{x} \right) g_{x}^{\left( k \right)}, \label{eq:cal_first_moment_v_x}\\
	m_{y}^{\left( k \right)}=\zeta _{y}m_{y}^{\left( k-1 \right)}+\left( 1-\zeta _{y} \right) g_{y}^{\left( k \right)}, \label{eq:cal_first_moment_v_y}
\end{subnumcases}
where $m_{x}^{\left( k \right)}$ and $m_{y}^{\left( k \right)}$ are the first moment (mean) of the gradient.
Secondly, the second step can be unfolded as 
\begin{subnumcases}
	~n_{x}^{\left( k \right)}=\varpi _{x}n_{x}^{\left( k-1 \right)}+\left( 1-\varpi _{x} \right) \left( g_{x}^{\left( k \right)} \right) ^2, \label{eq:cal_second_moment_v_x}\\
	n_{y}^{\left( k \right)}=\varpi _{y}n_{y}^{\left( k-1 \right)}+\left( 1-\varpi _{y} \right) \left( g_{y}^{\left( k \right)} \right) ^2, \label{eq:cal_second_moment_v_y}
\end{subnumcases}
where $n_x^{\left( k \right)}$ and $n_y^{\left( k \right)}$ are the second moment (uncentered variance) of the gradient.
Then, to counteract the biased moment estimations imposed by moving average, Adam introduces a correction step, which is specified by the following equations:
\begin{align}
    \begin{cases}
	\hat{m}_{x}^{\left( k \right)}=m_{x}^{\left( k \right)}/\left( 1-\zeta _{x}^{k} \right),  \label{eq:corrector_v_x} \\
	\hat{n}_{x}^{\left( k \right)}=n_{x}^{\left( k \right)}/\left( 1-\varpi _{x}^{k} \right), 
\end{cases}
\end{align}
\begin{align}
\begin{cases}
\label{eq:corrector_v_y}
	\hat{m}_{y}^{\left( k \right)}=m_{y}^{\left( k \right)}/\left( 1-\zeta _{y}^{k} \right) , \\
	\hat{n}_{y}^{\left( k \right)}=n_{y}^{\left( k \right)}/\left( 1-\varpi _{y}^{k} \right) .
\end{cases}
\end{align}
Finally, in the last step, the velocities are updated by taking a step toward the searching direction, i.e., 
\begin{subnumcases}
	~v_{k}^{(x)}=v_{k-1}^{(x)}+\frac{\alpha _{x}\hat{m}_{x}^{\left( k \right)}}{\sqrt{\hat{n}_{x}^{\left( k \right)}+\epsilon}}, \label{eq:update_v_x} \\
	v_{k}^{(y)}=v_{k-1}^{(y)}+\frac{\alpha _{y}\hat{m}_{y}^{\left( k \right)}}{\sqrt{\hat{n}_{y}^{\left( k \right)}+\epsilon}},  \label{eq:update_v_y}
\end{subnumcases}
where $\epsilon$ is a small real value to guarantee the numerical stability of this step.
The overall algorithm is summarized in \textbf{Algorithm \ref{alg:AGD-AO}} and illustrated in Fig. \ref{fig:AGD-AO}.

\begin{algorithm}[t!]
    \SetAlgoLined 
	\caption{Predictive Beamforming via AGD-AO}\label{alg:AGD-AO}
	\KwIn{Learning rate $\alpha _{x}$ and $\alpha _{y}$, $\epsilon$, maximum iteration $K$, stop criteria $\Gamma_{v_y}$ and $\Gamma_{v_y}$, previous estimated position $\hat{\mathbf{p}}_{l-1}$, and previous estimated velocity $\hat{\mathbf{v}}_{l-1}$. } 
	\KwOut{Beamformers for the next CPI, i.e., $\mathcal{F} _{l+1}$ }
    Calculate the current position of the user via $\hat{\mathbf{p}}_{l} = \hat{\mathbf{p}}_{l-1}+\Delta T\hat{\mathbf{v}}_{l-1}$, obtain the initial value for velocity estimation via $\mathbf{v} = \hat{\mathbf{v}}_{l-1}$, and let $k=1$ \;
    \While{$k < K$ and $|(v_{k}^{(x)}-v_{k-1}^{(x)}) / v_{k}^{(x)}| < \Gamma_{v_x}$ and $|(v_{k}^{(y)}-v_{k-1}^{(y)}) / v_{k}^{(y)}| < \Gamma_{v_y}$}{
    \tcp{Optimizing $v^{(x)}$}
    Calculate gradient $g_{x}^{\left( k \right)}$ using \eqref{eq:grad_v_x}\;
    Calculate the first moment $m_{x}^{\left( k \right)}$ using \eqref{eq:cal_first_moment_v_x}\;
    Calculate the second moment $n_{x}^{\left( k \right)}$ using \eqref{eq:cal_second_moment_v_x}\;
    Update the moment estimation and obtain $\hat{m}_{x}^{\left( k \right)}$ and $\hat{n}_{x}^{\left( k \right)}$ using \eqref{eq:corrector_v_x}
    Update velocity $v_{k}^{(x)}$ using \eqref{eq:cal_second_moment_v_x}\;
    
    \tcp{Optimizing $v^{(y)}$}
        Calculate gradient $g_{y}^{\left( k \right)}$ using \eqref{eq:grad_v_y}\;
        Calculate the first moment $m_{y}^{\left( k \right)}$ using \eqref{eq:cal_first_moment_v_y}\;
        Calculate the second moment $n_{y}^{\left( k \right)}$ using \eqref{eq:cal_second_moment_v_y}\;
        Update the moment estimation and obtain $\hat{m}_{y}^{\left( k \right)}$ and $\hat{n}_{y}^{\left( k \right)}$ using \eqref{eq:corrector_v_x}\;
        Update velocity $v_{k}^{(y)}$ using \eqref{eq:cal_second_moment_v_y}\;
        Enter the next iteration via $k = k+1$\;
    }
    Design predictive beamformer via \textbf{Algorithm \ref{alg:beamformer_design}} with $\hat{\mathbf{p}}_{l}$ and $\hat{\mathbf{v}}_{l}=[v_{k}^{(x)}, v_{k}^{(y)}]$.
 \end{algorithm}

\subsection{Computational Complexity and Stability Analysis}
\subsubsection{Computational Complexity Analysis}
The computational complexity of AGD-AO stems from three aspects, i.e., gradient calculation, moment updates, and parameter updates.
Here, we denote the dimension of the motion state and the observation vector as $U$ and $M$, respectively.
Thus, the complexity of gradient calculation of the velocities can be specified by $\mathcal{O}(UM)$.
Then, the complexity for moment updates is $\mathcal{O}(U)$.
Correspondingly, the complexity for parameter updates is $\mathcal{O}(U)$.
Finally, considering there will be $K$ iteration in the worst case, the total computational complexity for AGD-AO at one CPI is given by $\mathcal{O}(K(UM+U))$.
Considering $M \gg U$, the computational complexity can be simplified to $\mathcal{O}(KUM)$.

\subsubsection{Stability Analysis}
The stability of this method highly relies on the channel condition.
Specifically, when the channel is perturbed by low received SNR, the optimization landscape will be adversely affected.
In this case, the direction of gradient descent might be unstable.
Besides, since AGD-AO only estimates the velocities, it may suffer from error accumulation issues.
This means that errors in velocity estimation will accumulate, thereby hampering trajectory tracking. 
\section{EKF Method for Predictive Beamforming} \label{sect:ekf_solution}
\begin{figure*}[ht!]
	\centering
	\includegraphics[width=0.95\linewidth]{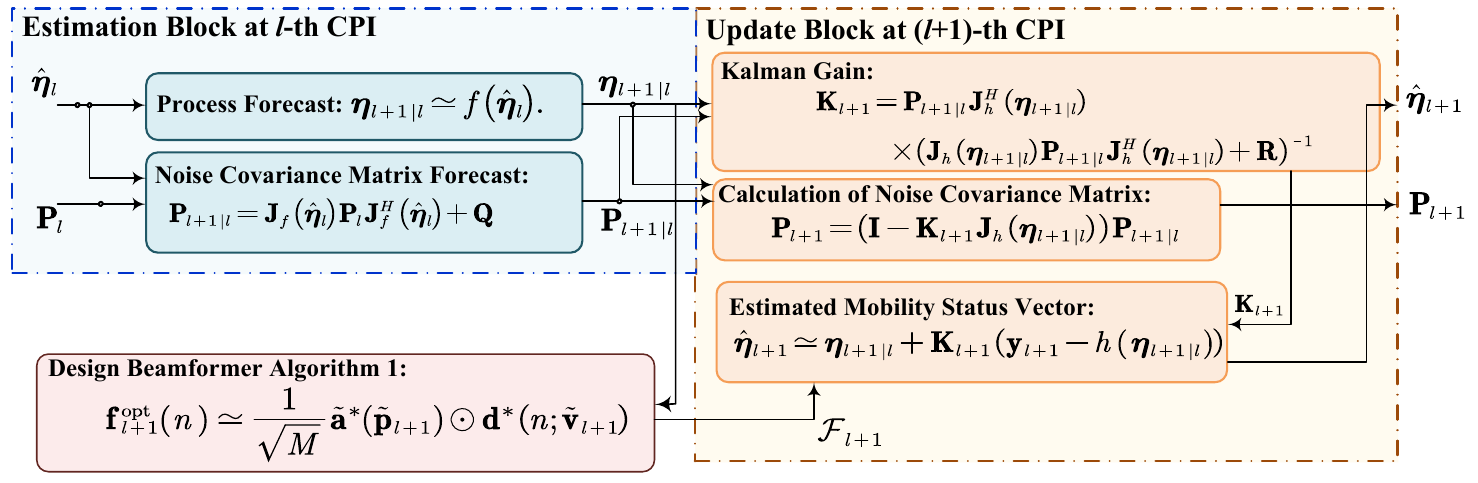}
	\caption{An overview of the EKF method. This process includes an estimation block and an update block.} \label{fig:algorithm_struct_EKF}
\end{figure*}
In this section, we propose an EKF method to solve this problem, which is a tailored version of a linear Kalman filter for non-linear observations and processes.  
The core idea of EKF is to linearize the estimation of the current mean and covariance using the Taylor series.
The observation and the kinematic function are specified by \eqref{eq:observation} and \eqref{eq:process}, with the covariance of $\mathbf{R}$ and $\mathbf{Q}$, respectively.
Since the rationale behind EKF has been well investigated, we will only present how to tailor EKF for our use, which contains three steps: 1)~initialization, 2)~model forecast step, and 3)~data assimilation step.
\subsection{Initialization}
For the initialization step $l=1$, we initialize the initial estimation $\hat{\boldsymbol{\eta }}_1=\boldsymbol{\eta }_1$ via IA and randomly initialize noise covariance matrix $\mathbf{P}_1$ as a diagonal matrix.   

\subsection{Model Forecast Step}
Since the kinematic function is known at the BS, the priori motion estimation $\boldsymbol{\eta }_{l+1|l} \in \mathbb{R}^{4 \times 1}$ can be obtained by forwarding the estimated user's motion state in the current CPI $\hat{\boldsymbol{\eta }}_{l}$ through the kinematic function, i.e.,
\begin{align}
	\boldsymbol{\eta }_{l+1|l}\simeq f\left( \hat{\boldsymbol{\eta }}_{l} \right). \label{eq:model-forecast}
\end{align}
It is noted that \eqref{eq:model-forecast} is not an accurate result since the accelerations contained in the motion state model are omitted.
We can also obtain the priori noise covariance matrix $\mathbf{P}_{l+1|l}$ by the following equation, i.e., 
\begin{align}
\mathbf{P}_{l+1|l}=\mathbf{J}_f\left(\hat{\boldsymbol{\eta }}_{l}\right) \mathbf{P}_{l}\mathbf{J}_{f}^{H}\left( \hat{\boldsymbol{\eta }}_{l} \right) +\mathbf{Q}, \label{eq:covariance_forecast}
\end{align}
where $\mathbf{J}_f\left( \hat{\boldsymbol{\eta }}_{l} \right)$ denotes the Jacobian matrix of the kinematic function $f(\cdot)$ evaluated at $\hat{\boldsymbol{\eta }}_{l}$.
Since the kinematic model is linear, the Jacobian matrix $\mathbf{J}_f\left( \cdot \right)$ can be expressed as  
\begin{align}
    \mathbf{J}_f\left( \cdot \right) =\left[ \begin{matrix}
	1&		0&		\bigtriangleup T&		0\\
	0&		1&		0&		\bigtriangleup T\\
	0&		0&		1&		0\\
	0&		0&		0&		1\\
    \end{matrix} \right].  
\end{align}
With the priori motion estimation $\boldsymbol{\eta }_{l+1|l}$, the predictive beamforming can be carried out with \textbf{Algorithm \ref{alg:beamformer_design}}.

\subsection{Data Assimilation Step}
After predictive beamforming at the $(l+1)$-th CPI, the observation will be received at the BS at the end of this CPI.
Based on this reception, a data assimilation step can be carried out to rectify the priori predictions $\boldsymbol{\eta }_{l+1|l}$ according to the posteriori observation $\mathbf{y}_{l+1}$.
According to the EKF theory, the posteriori estimation $\hat{\boldsymbol{\eta }}_{l+1}$ can be updated according to what follows:
\begin{align}
	\hat{\boldsymbol{\eta }}_{l+1}\simeq \boldsymbol{\eta }_{l+1|l}+\mathbf{K}_{l+1}\left( \mathbf{y}_{l+1}-h( \boldsymbol{\eta }_{l+1|l}) \right), \label{eq:estimation_update}
\end{align}
where the Kalman gain for the $(l+1)$-th CPI, denoted by $\mathbf{K}_{l+1}$, can be calculated by
\begin{align}
\mathbf{K}_{l+1}&=\mathbf{P}_{l+1|l}\mathbf{J}_{h}^{H}\left( \boldsymbol{\eta }_{l+1|l} \right) \notag \\&\quad \times\left( \mathbf{J}_{h}^{}\left( \boldsymbol{\eta }_{l+1|l} \right) \mathbf{P}_{l+1|l}\mathbf{J}_{h}^{H}\left( \boldsymbol{\eta }_{l+1|l} \right) +\mathbf{R} \right) ^{-1}, \label{eq:calculate_Kalman_gain}
\end{align}
in which $\mathbf{J}_{h}\left( {\boldsymbol{\eta}} \right) =\left[ \boldsymbol{\varphi}^{\left( 1 \right)},\boldsymbol{\varphi}^{\left( 2 \right)},\boldsymbol{\varphi}^{\left( 3 \right)},\boldsymbol{\varphi}^{\left( 4 \right)} \right] $ denotes the Jacobian of the measurement vector $h(\cdot)$ with respect to $\boldsymbol{\eta}$.
For brevity, the derivation of $\mathbf{J}_{h}(\cdot)$ is elaborated in Appendix \ref{appendix:A}.
Then, the posteriori noise covariance matrix can be calculated by
\begin{align}
    \mathbf{P}_{l+1}&=\left( \mathbf{I}-\mathbf{K}_{l+1}\mathbf{J}_{h}^{}\left( \boldsymbol{\eta }_{l+1|l} \right) \right) \mathbf{P}_{l+1|l}. \label{eq:update_covariance}
\end{align}
\begin{figure*}[ht!]
	\centering
	\subfloat[Optimization process of $v^{(x)}$]{
		\includegraphics[width=0.32\linewidth]{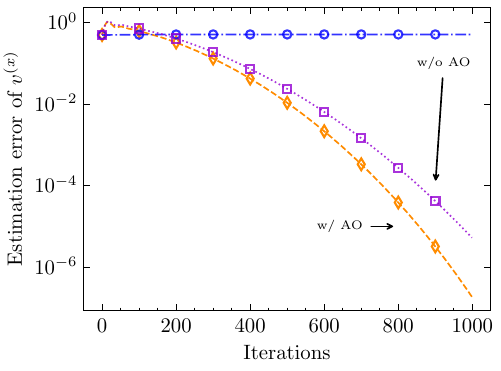}}
	\subfloat[Optimization process of $v^{(y)}$]{
		\includegraphics[width=0.32\linewidth]{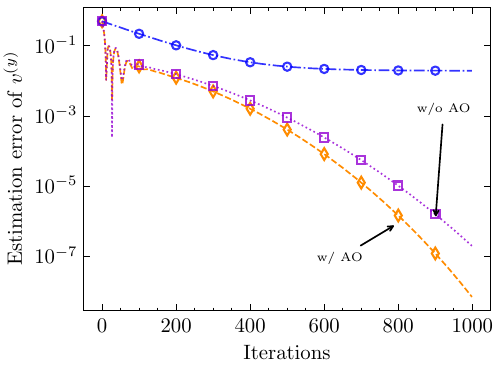}}
        \subfloat[The value of the objective function]{
		\includegraphics[width=0.32\linewidth]{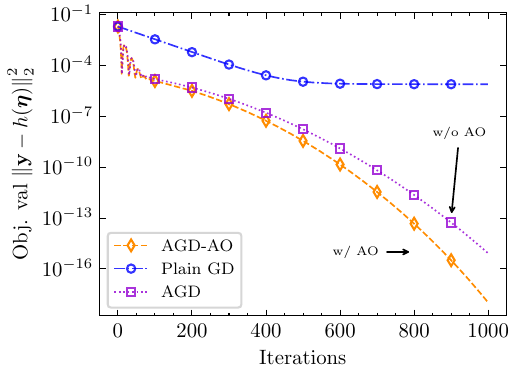}}
	\caption{A comparison of the convergence behavior of different algorithms.} \label{fig:beamfocosing_performance}\label{fig:algorithm_comparision}  
\end{figure*}
Finally, the structure of the proposed method is illustrated in Fig. \ref{fig:algorithm_struct_EKF}.
Additionally, the complete algorithm is summarized in \textbf{Algorithm \ref{alg:ekf}}.
\begin{algorithm}[t!]
    \SetAlgoLined 
	\caption{Predictive Beamforming via EKF}\label{alg:ekf}
	\KwIn{Estimated user's position and velocity for the current $l$-th CPI, i.e., $\hat{\mathbf{p}}_{l}$ and $\hat{\mathbf{v}}_{l}$, and the noise covariance matrix $\mathbf{P}_l$. } 
	\KwOut{Beamformers for the next CPI, i.e., $\mathcal{F} _{l+1}$}
	\tcp{Model Forecast Step:}\
	Obtain $\boldsymbol{\eta }_{l+1|l}$ and $\mathbf{P}_{l+1|l}$ by forecasting $\hat{\boldsymbol{\eta}}_{l}$ and $\mathbf{P}_{l}$ through \eqref{eq:model-forecast} and \eqref{eq:covariance_forecast}, respectively\;
 \tcp{Predictive Beamforming Step:}\
 Design predictive beamformer via \textbf{Algorithm \ref{alg:beamformer_design}} with $\boldsymbol{\eta }_{l+1|l}$\;
 \tcp{Data Assimilation Step:}\
    Obtain Kalman gain $\mathbf{K}_{l+1}$ via \eqref{eq:calculate_Kalman_gain}\;
    Obtain the posteriori user's motion state $\hat{\boldsymbol{\eta }}_{l+1}$ via \eqref{eq:estimation_update}.\;
    Obtain the posteriori noise covariance matrix $\mathbf{P}_{l+1}$ using \eqref{eq:update_covariance}\;
\Return{The posteriori motion state $\hat{\boldsymbol{\eta}}_{l+1}$ and the posteriori noise covariance matrix $\mathbf{P}_{l+1}$.}
\end{algorithm}

\subsection{Computational Complexity and Stability Analysis}
\subsubsection{Computational Complexity Analysis}
The computational complexity of EKF arises from two main aspects, i.e., the model forecast step and the update step.
Given the dimension of the user's motion state $U$, the complexity for the model forecast step can be written as $\mathcal{O}(U^3 + U^2)$, primarily due to the covariance forecast step.
Hence, the complexity of gradient calculation is $\mathcal{O}(MU + U^2)$.
For the data assimilation step, the computational complexity of the update step is as follows: $\mathcal{O}(M^2U + MU^2+M^3)$ for the calculation of Kalman gain, $\mathcal{O}(MU)$ for the state update, and $\mathcal{O}(M^2U+MU^2)$ for the covariance update.
Therefore, the overall computational complexity of the EKF method is given by $\mathcal{O}(M^3 + MU^2 + M^2U + MU +U^3+U^2)$.
Considering $M \gg U$, the computational complexity can be simplified as $\mathcal{O}(M^3)$.

\subsubsection{Stability Analysis}
Similar to the AGD-AO method, the stability of the EKF method highly relies on the channel condition. 
However, unlike the AGD-AO method, the EKF incorporates the observations from multiple CPIs. 
Consequently, the EKF's tracking is in a closed-loop control, which makes the EKF method more robust to variations in channel conditions.

\section{Simulation results} \label{sect:simulation_res}
In this section, the simulation results are presented to verify the effectiveness of the proposed AGD-AO and EKF approaches.
The following parameter setups are utilized throughout our simulations unless otherwise specified.

\subsection{Simulation Setups}
For physical-layer parameters, it is assumed that the BS is equipped with $M=512$ antennas with half-wavelength spacing and transmits signals downlink to a single-antenna user.
The carrier frequency and bandwidth of the communication network are set to $f=30$ GHz and $W=100$ kHz, respectively.
Consequently, the symbol duration is set to $10^{-5}$ s, and the time duration of one CPI is set to $\Delta T = 10^{-4}$ s, indicating that $N=10$.
The pathloss model for the communication signal is specified by $\alpha_1(\mathbf{p})=\beta (\sqrt{x^2+y^2})^{-2}$ \cite{liu2020radar}, while $\beta$ denotes the channel power gain at the reference point.
Additionally, according to \cite{liu2020radar}, the pathloss model for the echo signal is given by $\alpha_2(\mathbf{p})=\sigma_{\rm RCS} \beta (2\sqrt{x^2+y^2})^{-2}$.
Since the Swerling I model is adopted in this work, we can set the $\sigma_{\rm RCS}$ as a constant number in line with many existing related works \cite{ yuan2021bayesian, mu2021integrated, liu2022learning, zhang2024transformer, zhang2023deep}, whose value can be obtained via IA.
Therefore, with loss of generality, we set $\beta =1$ and $\sigma_{\rm RCS}=1$.
The transmit power at the BS is set to $P=30$ dBm, with the noise power set to $\sigma_{c}^2 =\sigma_e^2= 10^{-8}$.
For the motion state model, we set the velocity variances within each CPI to $\sigma_{v^{(x)}}^2=0.01$ and $\sigma_{v^{(y)}}^2=0.01$ for $v^{(x)}$ and $v^{(y)}$, respectively.
The initial coordinates of the user are set to $x_0 = 5~\mathrm{m}$ and $y_0 = 10~\mathrm{m}$, while the initial velocities of the user are set to $v_0^{(x)}=8~m/s$ and $v_0^{(y)}=7~m/s$.
The total number of CPI is $20000$ or equivalently 20 s.
For the algorithm's parameters, we set AGD-AO's learning rate and iteration threshold to $\alpha _{x}= \alpha _{y}= 5 \times 10^{-2}$ and  $\Gamma_{v_x}=\Gamma_{v_y}=10^{-5}$, respectively,
While other hyper-parameters are set to $\epsilon=10^{-8}$, $\zeta _{x}= \zeta _{y}=0.9$, $\varpi _{x}=\varpi _{y}=0.999$, and $K=500$.
In the EKF method, we set $\mathbf{P}_1=0.1 \times \mathbf{1}_4$.

\subsection{Effectiveness of the AGD-AO Method}
To verify the effectiveness of the proposed AGD-AO algorithm, we first compare it with two benchmark algorithms:
\begin{itemize}
    \item \textbf{Plain gradient descent (GD)}:~A plain method performs gradient descent directly without using moment estimation.
    \item \textbf{AGD}:~The framework of this method is the same as the proposed AGD-AO method, with the only difference being the absence of the alternative optimization structure.
    Therefore, the velocities $v^{(x)}$ and $v^{(y)}$ are optimized simultaneously.
\end{itemize}
As illustrated in Fig. \ref{fig:algorithm_comparision}, we compare the rate of convergence between the proposed AGD-AO method and other benchmarks.
The performance is evaluated using root square error to the ground-truth (GT) position and velocities, i.e., $\boldsymbol{\eta}=[0~m, 10~m, 8~m/s, 7~m/s]$.
It can be seen from this figure that, compared to the plain GD method, both the proposed AGD-AO and AGD methods converge at a higher speed.
The poor performance of plain GD may be attributed to its inability to adaptively adjust the learning rate.
Due to the small value of $\Delta T$, the gradients of the objective function \eqref{eq:obj-AGD-AO} are small around the optimum, thus making it difficult for plain GD to find the direction of descending. 
In comparison, by taking the exponential average over history gradients (moment estimation) to automatically tune the learning rate, the AGD-AO and AGD methods can converge to the optimum at a faster pace.
Additionally, compared to the method proposed by \cite{wang2024near}, our method can save memory and reduce the computational complexity since no Hessian calculation is needed in the process.
Moreover, compared to AGD, AGD-AO can stabilize the optimization process by alternatively optimizing the velocities, enabling it to converge more quickly than its counterpart.

\subsection{Trajectory and Velocity Tracking Performance}
\begin{figure}[t!]
	\centering
	\includegraphics[width=0.9
 \linewidth]{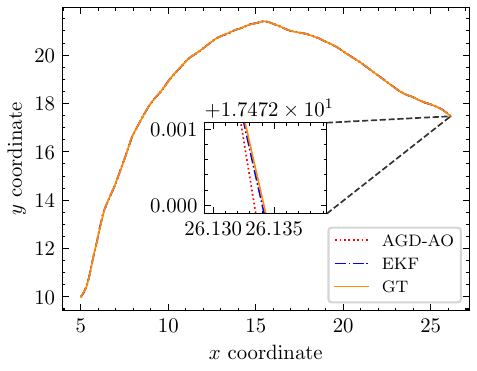} 
	\caption{An illustration of trajectory tracking performance of AGD-AO and EKF under $P=30$ dBm.} \label{fig:trajectory}
\end{figure}
In Fig. \ref{fig:trajectory}, we illustrate the user-tracking performance of algorithms under $P=30$ dBm.
Compared to far-field scenarios, the NISE-enabled user tracking/estimation scheme does not require the user to follow a certain trajectory.
Thus, NISE is capable of capturing the full motion state of the user, which is evidenced by the curved trajectory in this figure.
We can also observe that the trajectory tracked by the EKF method is closer to the ground truth (GT).
The reason lies in the fact that the EKF method utilizes multiple CPIs to track the user's trajectory.
However, it is worth noting that the difference in trajectory prediction performance between AGD-AO and EKF is still small.

\begin{figure*}[ht]
	\centering
	\subfloat[$v^{(x)}$ tracking performance.]{
		\includegraphics[width=0.48\linewidth]{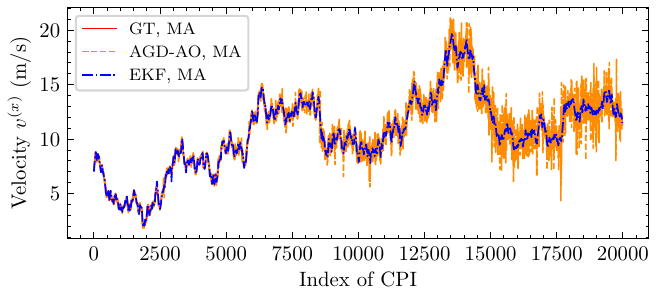} \label{fig:velocity_a}} 
	\subfloat[$v^{(y)}$ tracking performance]{
		\includegraphics[width=0.48\linewidth]{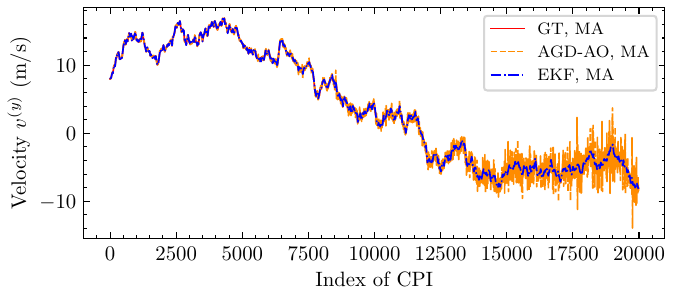}\label{fig:velocity_b}} \\
  \subfloat[$v^{(x)}$ error tracking]{
		\includegraphics[width=0.48\linewidth]{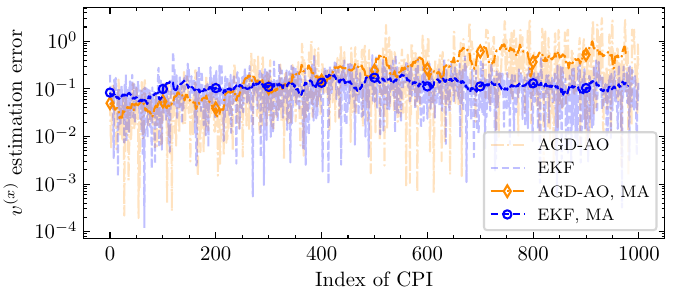}\label{fig:velocity_c}} 
  \subfloat[$v^{(y)}$ error tracking]{
		\includegraphics[width=0.48\linewidth]{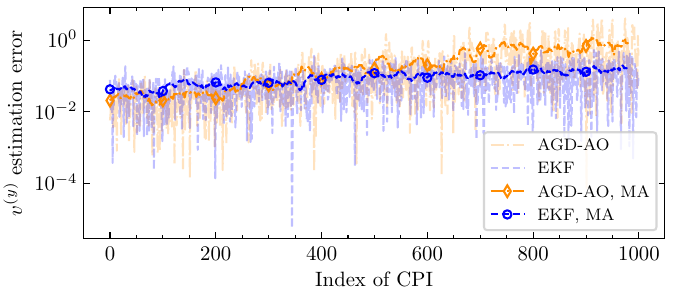}\label{fig:velocity_d}} 
	\caption{An illustration of velocity prediction performance under $P=30$ dBm.} \label{fig:velocity_prediction}
\end{figure*}
In Fig. \ref{fig:velocity_prediction}, we further investigate the velocity tracking performance of the proposed methods when the user follows the same trajectory illustrated in Fig. \ref{fig:trajectory}.
To better present the pattern hidden in the process, moving average (MA) with a window size of $20$ is utilized to smooth out the short-term fluctuations and highlight the long-term patterns, which is often used in time series analysis.
Specifically, the MA operation is defined as $y_l = (1/K) \sum_{k=1}^{K}x_{l-k+1} $, where $y$ denotes the index of the MA results, $x$ represents the datapoints, and $K$ denotes the MA window size.
It can be observed from Figs \ref{fig:velocity_a} and \ref{fig:velocity_b} that both the EKF and AGD-AO algorithms can follow the random evolution of velocities.
However, in contrast to the stable behavior of the EKF curve, the behavior of the AGD-AO algorithm is more random, which can be ascribed to the multiple CPIs that EKF employed.
Unlike the EKF method, the AGD-AO algorithm utilizes just one single CPI as its observation and performs estimation for each CPI independently.
In addition, as time evolves, we can see that the estimation error of the AGD-AO algorithm increases.
The reason lies in the fact that, as time progresses, the distance between the BS and the user increases accordingly.
Therefore, the gradients of \eqref{eq:obj-AGD-AO} are smaller, thereby causing more difficulties for gradient descending.
However, due to the incorporation of multiple CPIs, the EKF method is more robust to these changes.
To fortify this argument, in Figs. \ref{fig:velocity_c} and \ref{fig:velocity_d}, we examine the estimation errors of $v^{(x)}$ and $v^{(y)}$ with respect to the ground truth value in a logarithmic scale.
The errors are calculated via $|x-\hat{x}|$, where $x$ and $\hat{x}$ denote the ground truth and estimated values, respectively.
As demonstrated in Figs. \ref{fig:velocity_c} and \ref{fig:velocity_d}, when the user moves away from the BS, the velocity estimation error of the AGD-AO algorithm increases correspondingly.
On the contrary, the velocity estimation of the EKF method is more robust to the changes in distances between the transceivers.
It is also interesting to see that the estimation accuracy of the AGD-AO method outperforms that of the EKF method at the beginning.
This is because a stricter threshold is set for the AGD-AO method.
However, due to its lesser robustness against distance, the estimation performance of the EKF surpasses that of the AGD-AO method over time. 

\begin{figure}[ht]
	\centering
	\includegraphics[width=0.9\linewidth]{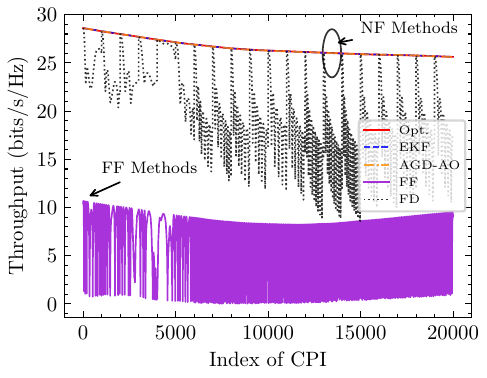} 
	\caption{An illustration of predictive beamforming performance, under $P=30$ dBm.} \label{fig:throughput}
\end{figure}
\subsection{Predictive Beamforming Performance}
In Fig. \ref{fig:throughput}, we study the predictive beamforming performance of the proposed methods compared to the following three benchmarks:
\begin{itemize}
    \item \textbf{FF}, which refers to the far-field (FF) method, utilizes real-time perfect channel state information (CSI) to perform beamforming design.
    However, unlike near-field beamforming, this method only considers the user's angle and radial velocity while neglecting the additional distance and transverse velocity.
    Therefore, the steering vector of FF can be expressed as 
    \begin{align}
        \mathbf{f}_l(n)&=\frac{1}{\sqrt{M}}e^{-j\frac{2\pi}{\lambda}\left(nT_sv^{(r)} \right)} \notag \\
        &\qquad \qquad \qquad  \times[e^{-j\frac{2\pi}{\lambda}\mathbf{p}^T\mathbf{k}_{1}^{}},...,e^{-j\frac{2\pi}{\lambda}\mathbf{p}^T\mathbf{k}_{M}^{}}]^T. \notag
    \end{align}
    \item \textbf{FD}, which refers to a feedback (FD)-based method, utilizes a feedback link to track the user.
    Specifically, the user reports its position and velocity every $0.1 s$ to the BS through the dedicated uplink feedback channel.
    \item \textbf{Opt.}, which refers to the optimal method, utilizes real-time CSI to carry out beamforming, which is the theoretical upper bound of channel capacity.
\end{itemize}

\begin{figure}[t!]
	\centering
	\includegraphics[width=0.9\linewidth]{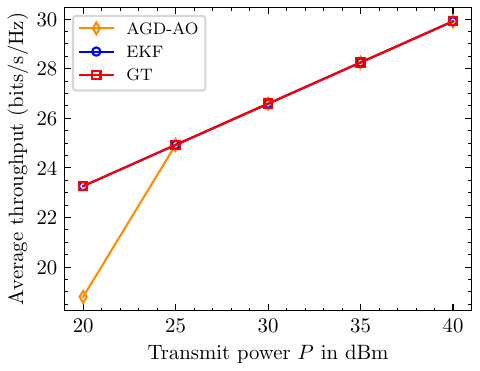}
	\caption{An illustration of average throughput under different transmit power budgets.} \label{fig:throughput_loss}
\end{figure}
As shown in Fig. \ref{fig:throughput}, the proposed methods can achieve optimal performance, which reinforces the effectiveness of our approaches.
Among the benchmarks, the FF scheme has the worst performance, which is attributed to its lack of distance and transverse velocity information.
It can be observed that the curve of the FF method exhibits a ``U" shape.
The downhill part of the curve can be justified by the fact that the increasing distance between the BS and the user will result in more server attenuation.
Then, the uphill part of the curve indicates that as the user moves away from the BS, the near-field channel vector is more similar to the far-field channel vector.
In this scenario, far-field beamsteering becomes effective, thus counteracting the attenuation introduced by pathloss.
For the FD scheme, its performance is not stable, due to its periodic feedback.
Although FD's performance can be improved with more frequent feedback, this would require a dedicated uplink feedback channel and increase communication overhead due to the additional uplink transmissions.

Fig. \ref{fig:throughput_loss} depicts the averaged throughput over all CPIs versus transmit power budgets.
As illustrated in this figure, the throughput grows as the transmit power increases.
In addition, in the high transmit power region, both of the schemes can strike near-optimal performance, indicating that the tracking performance is decent.
However, as the transmit power is tuned lower, the AGD-AO method experiences a degradation in averaged throughput, which reflects a decline in its tracking accuracy.
This degradation occurs due to the fact that, in the AGD-AO algorithm, only the velocities are estimated using the gradient-based method.
Consequently, the performance of velocity estimation determines the performance of trajectory tracking.
In this case, due to the interference introduced by noise, the optimization landscape is perturbed, thus misleading the gradient descent process.
In contrast, the EKF method keeps using information contained in multiple CPIs to rectify its tracking results, thus leading to a more robust performance.

\section{Conclusion}
\label{sect:conclu}
A NISE enabled predictive beamforming scheme has been proposed in this work.
Specifically, with the spherical wavefronts in the near-field region, the full motion state of the user, including the angle, distance, radial velocity, and transverse velocity, can be captured.
Based on the obtained full motion state, predictive beamforming can be carried out without the need for dedicated feedback links between transceivers.
A pair of methods is proposed to estimate and track the user's trajectory.
Specifically, when a single CPI is utilized, the gradient-based AGD-AO method exploits the moment estimation to find the user's velocity quickly, while the user's position can be obtained via the kinematic function.
On the contrary, when multiple CPIs are available at the BS, the EKF method can track the full motion state by updating its current estimation of the mean and noise covariance of the motion state model.
Simulation results fortified the effectiveness of both of the algorithms compared with FD and FF methods and further revealed that, with multiple CPIs at hand, the EKF method can achieve robust user tracking compared to the AGD-AO method.

\appendix
\section*{Derivation of Jacobian of Observation Function} 
\label{appendix:A}
In this Appendix, we omit the index of CPI to make the nations more succinct.
The Jacobian matrix of the observation function $h(\boldsymbol{\eta})$ with respect to the motion state vector $\boldsymbol{\eta}$ can be written as 
\begin{align}
    \mathbf{J}_h\left( \boldsymbol{\eta } \right)&=\frac{\partial h\left( \boldsymbol{\eta } \right)}{\partial \boldsymbol{\eta }}=\left[ \frac{\partial h\left( \boldsymbol{\eta } \right)}{\partial x},\frac{\partial h\left( \boldsymbol{\eta } \right)}{\partial y},\frac{\partial h\left( \boldsymbol{\eta } \right)}{\partial v^{(x)}},\frac{\partial h\left( \boldsymbol{\eta }\right)}{\partial v^{(y)}} \right] \notag \\
    &=\left[ \boldsymbol{\varphi}^{\left( 1 \right)},\boldsymbol{\varphi}^{\left( 2 \right)},\boldsymbol{\varphi}^{\left( 3 \right)},\boldsymbol{\varphi}^{\left( 4 \right)} \right] . \tag{A-1}\label{eq:A-1}
\end{align}
Due to the symmetry between $x$ and $y$, the derivations of $\boldsymbol{\varphi}^{\left( 1 \right)}$ and $\boldsymbol{\varphi}^{\left( 2 \right)}$ will be similar.
For brevity, we will provide the detailed derivations of $\boldsymbol{\varphi}^{\left( 1 \right)}$, while noting that $\boldsymbol{\varphi}^{\left( 2 \right)}$ can be obtained with minor changes.
Given that the observation vector at the last symbol time is of interest, we denote $\mathbf{a}=\mathbf{a}(N;\mathbf{v},\mathbf{p})$.
Moreover, with $\alpha _2 = \alpha _2(\mathbf{p}) 
$, $\boldsymbol{\varphi}^{\left( 1 \right)}$ can be expressed as
\begin{align}
    \boldsymbol{\varphi}^{\left( 1 \right)}&=\frac{\partial \alpha _2 }{\partial x}\mathbf{a}\mathbf{a}^{T}\mathbf{f}+\alpha _2\frac{\partial \left( \mathbf{a}^{}\mathbf{a}^{T}\mathbf{f} \right)}{\partial x} \notag \\ &=\mathrm{I}_1\mathrm{I}_2+\mathrm{I}_3\mathrm{I}_4, \notag
\end{align}
in which $\mathrm{I}_1=\frac{\partial \alpha _2}{\partial x}$ and $\mathrm{I}_3=\alpha _2$ need to be calculated according to the specific pathloss model, and the remaining terms can be derived by the following equations:
\begin{align}
\mathrm{I}_2 &= \mathbf{a}^{}\mathbf{a}^{T}\mathbf{f} \notag \\
&=\left[ \left[ \mathbf{a} \right] _1\sum\nolimits_{m=1}^M{\left[ \mathbf{a}^{} \right] _m\left[ \mathbf{f} \right] _m,...,\left[ \mathbf{a}^{} \right] _M\sum\nolimits_{m=1}^M{\left[ \mathbf{a}^{} \right] _m\left[ \mathbf{f} \right] _m}} \right] ^T,  \tag{A-2} \label{eq:a-2}\\
\mathrm{I}_4 &=\frac{\partial \left( \mathbf{a}^{}\mathbf{a}^{T}\mathbf{f} \right)}{\partial x}. \tag{A-3}
\end{align}
According to \eqref{eq:a-2}, each entry of $\mathrm{I}_4$ can be expressed as 
\begin{align}
	\left[ \frac{\partial \left( \mathbf{a}^{}\mathbf{a}^{T}\mathbf{f} \right)}{\partial x} \right] _m=\left[ \frac{\partial \mathbf{a}}{\partial x} \right] _m\sum\nolimits_{m=1}^M{\left[ \mathbf{a} \right] _m\left[ \mathbf{f} \right] _m}+\left[ \mathbf{a} \right] _m\frac{\partial \mathbf{a}^{T}}{\partial x}\mathbf{f}. \notag
\end{align}
By denoting $\mathbf{d}=\mathbf{d}\left( N;\mathbf{v} \right) $ and $\tilde{\mathbf{a}}=\tilde{\mathbf{a}}(\mathbf{p})$, we have
\begin{align}
    \frac{\partial \mathbf{a}^{}}{\partial x} &=\frac{\partial \left( \mathbf{d}\odot \tilde{\mathbf{a}} \right)}{\partial x}= \tilde{\mathbf{a}}\odot\frac{\partial \mathbf{d}}{\partial {x}}+\mathbf{d}\odot\frac{\partial \tilde{\mathbf{a}}}{\partial {x}}, \tag{A-4} \label{eq:partial_a_c}
\end{align} 
where we denote $\mathrm{I}_5\triangleq \frac{\partial \mathbf{d}}{\partial x}$ and $\mathrm{I}_6\triangleq \frac{\partial \tilde{\mathbf{a}}}{\partial x}$.
By defining $\mathbf{g} \triangleq \left[ g_{1},g_{2},...,g_{M} \right] ^T\in \mathbb{R}^{M\times 1}
$ and $\mathbf{q}\triangleq \left[ q_{1},q_{2},...,q_{M} \right] ^T\in \mathbb{R}^{M\times 1}$, $\mathrm{I}_5$ can be derived by
\begin{align}
	\mathrm{I}_5\triangleq \frac{\partial \mathbf{d}}{\partial x}=-j\frac{2\pi}{\lambda}\Delta T\mathbf{d}\odot \left( v^{\left( x \right)}\frac{\partial \mathbf{g}}{\partial x}+v^{\left( y \right)}\frac{\partial \mathbf{q}}{\partial x} \right). \tag{A-5}\label{eq:mobility_model}
\end{align}
Then, since the coordinates of the $m$-th antenna of the ULA at the BS is denoted by $\mathbf{k}_m$, we also have $r _{m}^2\triangleq r _{m}^2 (\mathbf{p})= ([\mathbf{k}_m]_1-x ) ^2+([\mathbf{k}_m]_2 - y)^2$.
Therefore, the partial derivatives can be computed according to
\begin{align}
	\frac{\partial \mathbf{g}}{\partial x} &=\left[ \frac{\partial g_{1}}{\partial x},\frac{\partial g_{2}}{\partial x},...,\frac{\partial g_{M}}{\partial x} \right] ^T, \notag \\
	\frac{\partial g_{m}}{\partial x}&=\left([\mathbf{k}_m]_1-x \right) \left( \frac{\left|[\mathbf{k}_m]_1-x \right|}{r _{m}^{3}}-\frac{1}{\left| [\mathbf{k}_m]_1 -x\right|r _{m}} \right). \notag
\end{align}
Similarly, for $\mathbf{q}_{m}$, we have
\begin{align}
	\frac{\partial \mathbf{q}}{\partial x}&=\left[ \frac{\partial q_{1}}{\partial x},\frac{\partial q_{2}}{\partial x},...,\frac{\partial q_{M}}{\partial x} \right] ^T, \notag \\
	\frac{\partial q_{m}}{\partial x}&=\frac{\left| [\mathbf{k}_m]_2 - y \right|\left( [\mathbf{k}_m]_1 -x \right)}{r_{m}^{3}}. \notag 
\end{align}
For $\mathrm{I}_6$, we have  
\begin{align}
    \mathrm{I}_6\triangleq \frac{\partial \tilde{\mathbf{a}}}{\partial x}= -j\frac{2\pi}{\lambda} \tilde{\mathbf{a}} \odot \mathrm{I}_7, \tag{A-6}
\end{align}
where $\mathrm{I}_7=[ \frac{x-[\mathbf{k}_1]_1}{r_{1}},...,\frac{x - [\mathbf{k}_M]_1}{r_{M}} ] ^T$.
Since $y$ is equivalent to $x$, one can readily obtain $\boldsymbol{\varphi}^{\left( 2 \right)}$ by substituting $x$ with $y$.
In addition, $[\mathbf{k}_m]_1$ needs to be replaced by $[\mathbf{k}_m]_2$ for $\forall~m\in\mathcal{M}$.

Then, we calculate the partial derivative with respect to the velocities, i.e., $\boldsymbol{\varphi }^{\left( 3 \right)}$ and $\boldsymbol{\varphi }^{\left( 4 \right)}$.
Again, we provide the detailed derivations of $\boldsymbol{\varphi }^{\left( 3 \right)}$, while $\boldsymbol{\varphi }^{\left( 4 \right)}$ can be deducted via a similar approach.
First, we have 
\begin{align}
    \boldsymbol{\varphi }^{\left( 3 \right)} =\alpha^{(2)}\frac{\partial \left( \mathbf{a}\mathbf{a}^{T}\mathbf{f} \right)}{\partial v^{\left( x \right)}}, \tag{A-7}\label{eq:partial_v_3}
\end{align}
in which each entry is given by 
\begin{align}
	\left[ \frac{\partial \left( \mathbf{a}\mathbf{a}^{T}\mathbf{f} \right)}{\partial v^{\left( x \right)}} \right] _m=\left[ \frac{\partial \mathbf{a}}{\partial v^{\left( x \right)}} \right] _m\sum\nolimits_{m=1}^M{\left[ \mathbf{a}\right] _m\left[ \mathbf{f} \right] _m}+\left[ \mathbf{a} \right] _m\frac{\partial \mathbf{a}^{T}}{\partial v^{\left( x \right)}}\mathbf{f}. \notag
\end{align}
Moreover, the partial derivative $\frac{\partial \mathbf{a}^{}}{\partial v^{\left( x \right)}}$ can be calculated by
\begin{align}
    \frac{\partial \mathbf{a}^{}}{\partial v^{\left( x \right)}}&=\frac{\partial \left( \mathbf{d}\odot \tilde{\mathbf{a}} \right)}{\partial v^{\left( x \right)}}=\tilde{\mathbf{a}}\odot \frac{\partial \mathbf{d}}{\partial v^{\left( x \right)}}, \\ 
    \frac{\partial \mathbf{d}}{\partial v^{(x)}}&=\left( -j\frac{2\pi}{\lambda} \Delta T \right) \mathbf{d}\odot \mathbf{g}. \notag
\end{align}
Using the same method, $\boldsymbol{\varphi}^{\left( 4 \right)}$ can be calculated by substituting $v^{\left( x \right)}$ with $v^{\left( y \right)}$. 
Specially, partial derivatives $\frac{\partial \mathbf{a}^{}}{\partial v^{\left( x \right)}}$ needs to be replaced by $\frac{\partial \mathbf{d}_c}{\partial v^{(y)}}=\left( -j\frac{2 \pi}{\lambda } \Delta T \right) \mathbf{d}\odot \mathbf{q}$.

\bibliographystyle{IEEEtran}
\bibliography{reference/mybib}
\end{document}